\documentclass[12pt, draftclsnofoot, onecolumn]{IEEEtran}
\usepackage{array}
\IEEEoverridecommandlockouts
\usepackage{cite}
\usepackage{amsmath,amssymb,amsfonts}
\usepackage{algpseudocode}
\usepackage{graphicx}
\usepackage{amsmath}
\usepackage{amsthm}
\usepackage{textcomp}
\usepackage{xcolor}
\usepackage{amsmath}
\usepackage{subfigure} 
\usepackage{algorithm}
\usepackage{algpseudocode}
\usepackage{subfigure}
\usepackage{lipsum}
\usepackage{color}

\newtheorem{lemma}{\emph{\underline{Lemma}}}

\def\phi{\varphi}

\def\l{\left}
\def\r{\right}
\def\({\left(}
\def\){\right)}

\def\b0{{\mathbf{0}}}

\newcommand{\nn}{\nonumber}

\begin{document}
\title{3D Placement for Multi-UAV Relaying: An Iterative Gibbs-Sampling and Block Coordinate Descent Optimization Approach}
        
\author{Zhenyu~Kang,~\IEEEmembership{Student~Member,~IEEE}        Changsheng~You,~\IEEEmembership{Member,~IEEE,}        and~Rui~Zhang,~\IEEEmembership{Fellow,~IEEE}}

\maketitle

\maketitle
\newcommand\blfootnote[1]{%
\begingroup
\renewcommand\thefootnote{}\footnote{#1}%
\addtocounter{footnote}{-1}%
\endgroup
}
\vspace{-22pt}
\begin{abstract}
In this paper, we consider an unmanned aerial vehicle (UAV) enabled relaying system where multiple UAVs are deployed as aerial relays to support simultaneous communications from a set of source nodes to their destination nodes on the ground. An optimization problem is formulated under practical channel models to maximize the minimum {achievable expected rate} among all pairs of ground nodes by jointly designing UAVs' three-dimensional (3D) placement as well as 
the bandwidth-and-power allocation. This problem, however, is non-convex and thus difficult to solve. As such, we propose a new method, called \textit{iterative Gibbs-sampling and block-coordinate-descent (IGS-BCD)}, to efficiently obtain a high-quality suboptimal solution by synergizing the advantages of both the deterministic (BCD) and stochastic (GS) optimization methods. 
Specifically, our proposed method alternates between two optimization phases until convergence is reached, namely, one phase that uses the BCD method to find locally-optimal UAVs’ 3D placement and the other phase that leverages the GS method to generate new UAVs' 3D placement for exploration. Moreover, we present an efficient method for properly initializing UAVs' placement that leads to faster convergence of the proposed IGS-BCD algorithm. Numerical results show that the proposed IGS-BCD and initialization methods outperform the conventional BCD or GS method alone in terms of convergence-and-performance trade-off, as well as other benchmark schemes.  

\end{abstract}
\blfootnote{Part of this work has been presented at the IEEE International Conference on Communications (ICC), Dublin, Ireland, June 2020  \cite{kang}.}
\blfootnote{The authors are with the Department of Electrical and Computer Engineering, National University of Singapore, Singapore 117583 (Email: zhenyu\_kang@u.nus.edu, eleyouc@nus.edu.sg, elezhang@nus.edu.sg).}

\vspace{-22pt}
\begin{IEEEkeywords}
UAV communication, aerial relay, 3D placement optimization, Gibbs sampling, block coordinate descent.
\end{IEEEkeywords}

\IEEEpeerreviewmaketitle

\section{Introduction}
Unmanned aerial vehicles (UAVs) are expected to be widely employed as new aerial communication platforms in future wireless networks to enhance the coverage and throughput of traditional terrestrial networks, by leveraging the advantages of UAVs including controllable maneuver, high mobility, flexible deployment as well as line-of-sight (LoS) dominant UAV-ground channels \cite{8918497}. This vision has spurred intensive enthusiasm in recent years to incorporate UAVs into wireless communication systems, leading to a variety of new applications, such as cellular-connected UAV \cite{b3,8624565,mei2019cellular}, UAV-assisted terrestrial communication \cite{8954798,b5,8675440}, UAV-enabled relaying \cite{7572068,8626132,8443133}, UAV-enabled wireless sensor networks \cite{you20193d,8515012,you2019hybrid}, and so on. 

Particularly, for high-mobility UAV-enabled relaying systems, UAV trajectory design has been extensively studied in the literature for e.g., maximizing the relaying communication throughput \cite{7572068} or UAV energy efficiency \cite{7888557} under the LoS channel condition.
Besides UAV trajectory optimization, another key design issue in UAV-enabled relaying is how to deploy quasi-static UAVs in the three-dimensional (3D) space for maximizing the communication rates of their aided ground nodes. An initial attempt for addressing this issue has been made in \cite{b6}, where the authors optimized UAVs' two-dimensional (2D) placement with fixed (minimum) altitude under the LoS channel model, which is reasonable for rural areas with UAV deployed at high altitude. 
However, such a simplified LoS UAV-ground channel model is practically inaccurate for urban areas with dense buildings, as it does not capture the non-negligible UAV-ground channel blockage, shadowing, and multi-path fading. 
As such, two more sophisticated channel models have been proposed to improve the accuracy. 
{\color{black} Specifically, for UAV deployed/flying at relatively low altitude, the shadowing due to obstacles (e.g., high-rise buildings) severely impairs the UAV-ground channels. To characterize it, a \textit{generalized probabilistic LoS} channel model for Manhattan-type cities was proposed in \cite{you2019hybrid}, where the LoS probability is modeled as a generalized logistic function of the UAV-ground node elevation angle.
This channel model was also adopted in \cite{8412127}, where UAVs are deployed to offload downlink data for maximizing the revenue of the ground macro base station.
In addition, the authors in \cite{8903530} proposed a nested segmented UAV-ground channel model and developed a low-complexity algorithm to search for the globally optimal UAV position by leveraging local terrain information.
On the other hand, for UAV at high altitude in urban areas, it has a high likelihood to establish LoS links with ground nodes and thus experiences less shadowing but non-negligible multi-path fading.} By using a data regression model-fitting approach, an \textit{elevation-angle dependent Rician fading} channel model was proposed in \cite{you20193d}, where the Rician factor in general increases with the UAV-ground node elevation angle due to less ground reflection and scattering. 
Intuitively, a larger elevation angle by moving the UAV horizontally closer to its served ground node and/or increasing its altitude above the ground result in less multi-path fading \cite{1469712}, while the higher altitude of UAV also leads to more path-loss due to the increasing UAV-ground distance. Thus, a major challenge in designing UAVs' 3D placement under the elevation-angle dependent Rician fading channel model is how to balance the aforementioned angle-distance trade-off for communication rate maximization, which has not been addressed in the literature to the authors' best knowledge.

It is worth noting that for UAV placement optimization, as the optimization problems are usually non-convex and difficult to solve, different approaches have been proposed in the literature for sub-optimally solving them, which can be roughly classified into two categories: \emph{deterministic} versus \emph{stochastic} UAV placement designs.
{\color{black} Among others, one typical deterministic method is the block coordinate descent (BCD) \cite{beck2013convergence}, which iteratively optimizes UAVs' placement and communication resource allocation.} This method has been widely utilized to e.g., optimize the placement of a single UAV for maximizing the number of covered users \cite{b6,b12}, and that of multiple UAVs for minimizing their total transmit power under a coverage constraint \cite{mozaffari2015drone}. Although the BCD method is computationally efficient, the obtained UAVs' placement may suffer considerable rate performance loss with a heuristically chosen UAVs' placement initialization since the converged solution is likely to trap in a low-quality local optimum. 
Besides, geometry-based methods, which leverage geometric features such as the locations of ground nodes, have also been used for designing UAV placement. For example, a dynamic clustering algorithm was proposed in \cite{8318674} to position the UAVs at the centroids of user clusters for saving their sum power consumption.
In \cite{7486987}, the circle packing method was utilized to maximize the coverage region of UAVs by adjusting the coverage areas of UAVs via their 3D locations.
Moreover, a space partition method was proposed in \cite{8712804}, \cite{8849271}, where the authors formulated a space quantization problem for designing UAVs' 2D placement and applied the Lloyd's algorithm to minimize the communication power consumption, by alternately partitioning the space into different small cells and updating their centroid points as UAVs' placement locations. 
However, these geometry-based methods, in general, cannot be applied when the communication requirement is a complicated function with respect to (w.r.t.) UAVs' placement under practical UAV-ground channel models. 
In contrast, stochastic methods usually leverage random simulations to generate UAV placement. For instance, a multi-population heuristic genetic algorithm (GA) was proposed in \cite{8600206} to maximize the number of covered users by generating random UAV placement following the natural selection process, i.e., selection, crossover, and mutation.
In addition, the particle swarm optimization (PSO) method was adopted in \cite{8735725} to maximize the communication throughput by modeling each UAV as an individual particle and adjusting UAVs' movement according to their utilities. This idea was, in fact, inherited from another stochastic optimization method, called \emph{Gibbs sampling} (GS), which has been applied to optimize the placement of base stations (BSs) on the ground for improving the throughput of a heterogeneous wireless cellular network \cite{6576466}. The key idea of GS lies in iteratively updating the state of each node according to well-designed transition probabilities so as to learn a near-optimal solution gradually. 
Nevertheless, these heuristic algorithms usually have slow convergence and may not necessarily have performance guarantee.
Moreover, it is worth mentioning that, although there have been some recent works that used reinforcement learning (RL) to design UAVs' placement, they mostly targeted to adaptively adjusting UAVs' placement according to the dynamic environment such as user movement \cite{8736350} instead of using RL for learning optimization solutions.
To summarize, the existing methods for UAV placement optimization usually suffer from either slow convergence (e.g., RL, GS) or considerable communication performance loss (e.g., BCD, clustering). This thus motivates this paper to design a new UAV placement optimization method for balancing the convergence-and-performance trade-off. 

For the purpose of exposition, we consider in this paper a multi-UAV relaying system where multiple UAVs are deployed to help relay data from a set of source nodes to their respective destination nodes on the ground, assuming that no direct link exists between any pair of ground nodes. Different from the existing works on UAV placement optimization that mostly adopted the simplified LoS channel model, we consider the practically more accurate elevation-angle dependent Rician fading UAV-ground channel model, under which we formulate an optimization problem to maximize the minimum {achievable expected rate} among all source-destination pairs subject to practical constraints on the transmit power of both the UAVs and source nodes, bandwidth, as well as the flow conservation for data relaying \cite{8443133}, i.e., for each data stream associated with a pair of source and destination nodes, a UAV forwards all the data that has been received from the source node and other UAVs. However, the optimal solution to this problem is difficult to obtain due to the coupling of transmit power, bandwidth, and UAVs' 3D placement in the achievable rate under the practical channel model, as well as the non-convex flow conservation constraint. As such, we propose a new method, called \textit{iterative GS and BCD} (IGS-BCD)\footnote{{\color{black}In this paper, we use the term of BCD for a maximization problem without causing confusion.}}, to efficiently obtain a high-quality suboptimal solution by synergizing the advantages of both the deterministic/BCD and stochastic/GS methods. Specifically, our proposed method alternates between two optimization phases, namely, a BCD phase that aims to quickly find locally-optimal UAVs' 3D placement and a GS phase that is designed for further improving the max-min rate by exploring new UAVs' locations. Moreover, in the BCD phase, we propose an efficient iterative algorithm to decouple the joint optimization into three sub-problems and iteratively solve them, namely, the optimizations of bandwidth-and-power allocation, UAVs’ horizontal placement, and UAVs’ vertical placement. Although these sub-problems are non-convex, we apply the successive convex approximation (SCA) technique to solve them sub-optimally. On the other hand, in the GS phase, we reformulate the max-min rate optimization problem into two sub-problems, corresponding to a slave problem for the bandwidth-and-power allocation optimization given fixed UAVs’ 3D placement and a master problem for UAVs’ 3D placement optimization. Although the slave problem can be efficiently solved by using the SCA technique, the master problem is intractable due to the lack of a closed-form expression for the max-min rate w.r.t. UAVs' 3D placement. To address this issue, we apply the GS method to gradually learn near-optimal UAVs' 3D placement by generating a sequence of samples for the UAVs' placement based on a Markov chain with the Markov transition probabilities determined by the max-min rates of different configurations of UAVs' placement. Furthermore, a high-quality UAVs' placement initialization method is proposed to accelerate the convergence speed of the proposed IGS-BCD algorithm. Numerical results show that our proposed IGS-BCD and initialization methods significantly improve the max-min rate with low complexity as compared to various benchmark schemes.

The remainder of this paper is organized as follows. 
The system model is introduced in Section~\ref{SM}, based on which, we formulate an optimization problem and present the main ideas of our proposed IGS-BCD algorithm in Section~\ref{PFPSM}.
The detailed designs for the BCD and GS phases of the proposed algorithm are elaborated in Section~\ref{bcd} and Section~\ref{gsb}, respectively. Simulation results and discussions are presented in Section~\ref{sr}. Finally, the conclusions are drawn in Section~\ref{clud}.

\vspace{-12.5pt}
\section{System Model}\label{SM}
\begin{figure}[t]
\centerline{\includegraphics[width=3in]{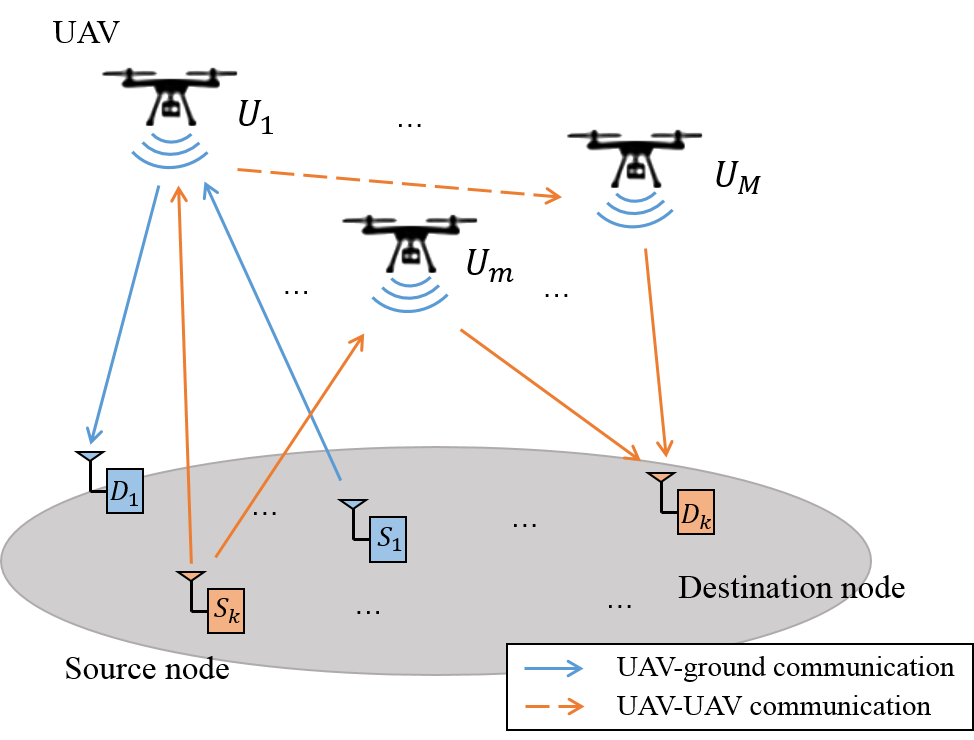}}
\caption{A multi-UAV relaying system for enabling communications between ground nodes.}
\label{sysmod}
\vspace{-12.5pt}
\end{figure}

\vspace{-4pt}
Consider a multi-UAV relaying system as illustrated in Fig. \ref{sysmod}, where $M$ UAVs, denoted by the set $\{U_m,m\in\mathcal{M}\}$ with $\mathcal{M}\triangleq\{1,\cdots,M\}$, are deployed as aerial relays to support simultaneous communications from $K$ ground source nodes to their respective ground destination nodes,
which are denoted by $\{S_k, k\in\mathcal{K}\}$ and $\{D_k, k\in\mathcal{K}\}$, respectively, with $\mathcal{K} \triangleq \{1, \cdots ,K\}$. 
Without loss of generality, let $[\boldsymbol{u}^{(\rm s)}_{k}, 0]$ and $[\boldsymbol{u}^{(\rm d)}_{k}, 0]$ denote respectively the 3D Cartesian coordinates of the $k$-th pair of source and destination nodes with $k\in\mathcal{K}$, where $\boldsymbol{u}^{(\rm s)}_{k}=[x^{(\rm s)}_{k}, y^{(\rm s)}_{k}]$ and $\boldsymbol{u}^{(\rm d)}_{k}=[x^{(\rm d)}_{k}, y^{(\rm d)}_{k}]$ are their corresponding horizontal coordinates\footnote{The superscript $(\rm s)$ and $(\rm d)$ represent the source and destination of the $k$-th pair ground nodes, respectively.}. 
Moreover, we assume that each pair of ground nodes are separated by a long distance such that the direct link between them is negligible due to severe terrestrial channel path-loss and blockage. To avoid obstacles such as buildings and conform to aerial regulations, the altitude of each UAV $U_m$ with $m\in\mathcal{M}$, denoted by $z_m$, is restricted within a range of $[H_{\min},H_{\max}]$. 
As such, the location of each UAV $U_m$ is represented by $[\boldsymbol{q}_m, z_m]$, where $\boldsymbol{q}_m\in \mathbb{R}^{1\times 2}$ denotes its horizontal coordinates.

\vspace{-12.5pt}
\subsection{Channel Model}
\vspace{-4pt}
Similar to \cite{you20193d}, we consider the Rician fading channel model for all the UAV-ground links, as UAV deployed at sufficiently high altitude has a high likelihood to establish LoS links with ground nodes, and at the same time, experiences small-scale fading due to ground scattering. As such,
the complex channel gain from each source node $S_k$ to UAV $U_m$ is modeled as
\begin{equation}\label{cpg1}
h_{k,m}^{(\mathrm{s})}=\sqrt{\beta _{k,m}^{(\mathrm{s})}} ~g_{k,m}^{(\mathrm{s})},
\end{equation}
where $\beta _{k,m}^{(\mathrm{s})}$ denotes the large-scale average channel power and $g_{k,m}^{(\mathrm{s})}$ denotes the small-scale fading coefficient. Specifically, let  $d_{k,m}^{(\mathrm{s})}$ denote the distance between source node $S_k$ and UAV $U_m$, which is given by
\begin{equation}\label{dkm}
d^{(\rm s)}_{k, m}=\sqrt{z_m^{2}+\|\boldsymbol{q}_{m}-\boldsymbol{u}^{(\rm s)}_{k}\|^{2}}.
\end{equation}
Then the average channel power gain, $\beta _{k,m}^{(\mathrm{s})}$, can be modeled as
\begin{equation}\label{beta_km}
\beta _{k,m}^{(\mathrm{s})}=\beta _0(d_{k,m}^{(\mathrm{s})})^{-\alpha},
\end{equation}
where $\alpha$ is the path-loss exponent that is usually in the range of $2\le\alpha\le 6$, $\beta_{0}$ is the channel power gain at the reference distance of $d_0=1$ meter (m).
On the other hand, the small-scale Rician fading can be modeled as
\begin{equation}
g_{k,m}^{(\mathrm{s})}=\sqrt{\frac{\kappa_{k,m}^{(\mathrm{s})}}{\kappa_{k,m}^{(\mathrm{s})}+1}}g+\sqrt{\frac{1}{\kappa_{k,m}^{(\mathrm{s})}+1}} \tilde{g},
\end{equation}
where $g$ corresponds to the LoS component with $|g|=1$, and $\tilde{g}$ denotes the random scattered Rayleigh fading component that is a zero-mean unit-variance circularly symmetric complex Gaussian (CSCG) random variable, $\kappa_{k,m}^{(\mathrm{s})}$ denotes the Rician factor of the channel from $S_k$ to $U_m$, which is the ratio between the power in the LoS component and fading component. According to \cite{1469712}, the Rician factor $\kappa_{k,m}^{(s)}$ can be modeled as the following function of the UAV-ground node elevation angle
\begin{equation}\label{hmk}
\kappa_{k,m}^{(\mathrm{s})}=A_{1} \exp (A_{2} \theta_{k,m}^{(\mathrm{s})}),
\end{equation}
\vspace{-4pt}
where $\theta_{k,m}^{(\mathrm{s})}=\arcsin (z_m / d^{(\rm s)}_{k, m})$, $A_1$ and $A_2$ are constants determined by the specific environment. Likewise, the channel gain from UAV $U_m$ to destination node $D_k$ can be modeled as 
$h_{m,k}^{(\mathrm{d})}=\sqrt{\beta_{m,k}^{(\mathrm{d})}} g_{m,k}^{(\mathrm{d})}$,
where the large-scale average channel power gain, $\beta_{m,k}^{(\mathrm{d})}$, and the small-scale fading coefficient, $g_{m,k}^{(\mathrm{d})}$, can be defined similar to $\beta_{k,m}^{(\mathrm{s})}$ and $g_{k,m}^{(\mathrm{s})}$, respectively.
For the UAV-UAV channels, due to the existence of LoS links between the UAVs, the channel gain from UAV $U_m$ to $U_n$ can be modeled as
\begin{equation}
h_{m, n}=\sqrt{\beta_{m,n}}{e}^{-j\tau_{m,n}}
=\frac{\sqrt{\beta_0}{e}^{-j\tau_{m,n}}}{\|\boldsymbol{q}_{m}-\boldsymbol{q}_{n}\|},
\end{equation}
where ${e}^{-j\tau_{m,n}}$ is the phase of $h_{m,n}$ due to the propagation delay from UAV $U_m$ to UAV $U_n$, i.e., $\tau_{m, n}=\frac{2 \pi d_{m, n}}{\lambda}$ with $\lambda$ denoting the carrier wavelength and $d_{ m, n}=\|\boldsymbol{q}_{m}-\boldsymbol{q}_{n}\|$.

\vspace{-12pt}
\subsection{Data Transmission Model}
\vspace{-4pt}
To avoid severe UAV-UAV and UAV-ground interference in the existence of LoS/LoS-dominant channels, we consider orthogonal transmissions for different communication links in separated frequency bands. 
Specifically, for the data transmission from each source node $S_k$ to UAV $U_m$, let $a_{k,m}^{(\mathrm{s})}\in[0,1]$ denote the allocated fraction of the total bandwidth, denoted by $B$ in Hertz (Hz), and $p_{k,m}^{(\mathrm{s})}$ denote the transmit power of $S_k$ in its allocated frequency band. Then the maximum {achievable expected rate} from $S_k$ to $U_m$, denoted by $C_{k,m}^{(\mathrm{s})}$  in bits per second per Hz (bps/Hz), is given by
\begin{equation}
C_{k,m}^{(\mathrm{s})}=a_{k,m}^{(\mathrm{s})}\log _{2}\left(1+\frac{|h_{k,m}^{(\mathrm{s})}|^{2} p_{k,m}^{(\mathrm{s})}}{a_{k,m}^{(\mathrm{s})}BN_0\Gamma  }\right),
\end{equation}
where $N_0$ denotes the power spectral density of the additive white Gaussian noise (AWGN) at the receiver, and $\Gamma>1$ denotes the gap of signal-to-noise ratio (SNR) between practical modulation-and-coding scheme and the theoretical Gaussian signaling. 

Let $R_{k,m}^{(\mathrm{s})}$ denote the fixed transmission rate from source node $S_k$ to UAV $U_m$. 
Then by assuming no CSI at the transmitters, the outage probability that UAV $U_m$ cannot successfully receive the data from source node $S_k$ can be expressed as
\begin{equation}\label{pkm}
\begin{aligned}
\mathcal{P}_{k,m}^{(\mathrm{s})}&=\mathbb{P}\left(C_{k,m}^{(\mathrm{s})}<R_{k,m}^{(\mathrm{s})}\right)\\&=\mathbb{P}\left(|g_{k,m}^{(\mathrm{s})}|^{2}<\frac{a_{k,m}^{(\mathrm{s})}BN_0 \Gamma(2^{R_{k,m}^{(\mathrm{s})}}-1)}{\beta_{k,m}^{(\mathrm{s})}p_{k,m}^{(\mathrm{s})}}\right) \\
&=F_{k,m}^{(\mathrm{s})}\left(\frac{a_{k,m}^{(\mathrm{s})}BN_0 \Gamma(2^{R_{k,m}^{(\mathrm{s})}}-1)}{\beta_{k,m}^{(\mathrm{s})} p_{k,m}^{(\mathrm{s})}}\right),
\end{aligned}
\end{equation}
where $F_{k,m}^{(\mathrm{s})}(u)$ is the non-decreasing cumulative distribution function (CDF) of the random variable $|g_{k,m}^{(\mathrm{s})}|^{2}$ w.r.t. $R_{k,m}^{(\mathrm{s})}$, and the CDF can be explicitly expressed as
\begin{equation}\label{Fkm}
F_{k,m}^{(\mathrm{s})}(u)=1-Q_{1}\l(\sqrt{2 \kappa_{k,m}^{(\mathrm{s})}}, \quad \sqrt{2(\kappa_{k,m}^{(\mathrm{s})}+1) u}\r),
\end{equation}
where $Q_1(x,y)$ denotes the standard Marcum-Q function \cite{marcum}. To ensure the transmitted data being reliably received as well as maximize the achievable rate, the transmission rate $R_{k,m}^{(\mathrm{s})}$ is chosen such that $\mathcal{P}_{k,m}^{(\mathrm{s})}=\epsilon_0, \forall k,m$, where $0<\epsilon_0\le 0.1$ is the maximum tolerable outage probability. Combining \eqref{beta_km} and \eqref{pkm} with $\mathcal{P}_{k,m}^{(\mathrm{s})}=\epsilon_0$ yields the maximum achievable expected (outage-aware) rate from source node $S_k$ to UAV $U_m$, $R_{k,m}^{(\mathrm{s})}$, which is given by
\begin{equation}
\begin{aligned}
R_{k,m}^{(\mathrm{s})}=a_{k,m}^{(\mathrm{s})}\log _{2}\!\left(\!1+\frac{\phi_{k,m}^{(\mathrm{s})}p_{k,m}^{(\mathrm{s})} \gamma_{0}}{(d_{k,m}^{(\mathrm{s})})^{\alpha / 2}a_{k,m}^{(\mathrm{s})}}\!\right)\! =a_{k,m}^{(\mathrm{s})}\log _{2}\!\left(\!1+\frac{\phi_{k,m}^{(\mathrm{s})}p_{k,m}^{(\mathrm{s})} \gamma_{0}}{({z_m^{2}+\|\boldsymbol{q}_{m}-\boldsymbol{u}^{(\rm s)}_{k}\|^{2}})^{\alpha / 2}a_{k,m}^{(\mathrm{s})}}\!\right)\!,
\end{aligned}
\end{equation}
where $\gamma_{0}\triangleq \frac{\beta_{0}}{N_{0}B\Gamma}$, and $\phi_{k,m}^{(\mathrm{s})}$ denotes the solution to $F_{k,m}^{(\mathrm{s})}(u)=\epsilon_0$. Although there is no closed-form expression for $\phi_{k,m}^{(\mathrm{s})}$ since it depends on the Rician factor, $\kappa_{k,m}^{(\mathrm{s})}$, which in turn depends on the UAV's 3D placement, $\{\boldsymbol{q}_m, z_m\}$ (see \eqref{dkm}--\eqref{hmk} and \eqref{pkm}--\eqref{Fkm}), it can be accurately approximated by the following logistic function \cite{you20193d}
\begin{equation}
\phi_{k,m}^{(\mathrm{s})}\approx f(v_{k,m}^{(\mathrm{s})})\triangleq C_{1}+\frac{C_{2}}{1+e^{-(B_{1}+B_{2} v_{k,m}^{(\mathrm{s})})}},
\end{equation} 
where the coefficients $B_1<0$, $B_2>0$, $C_1 > 0$ and $C_2 > 0$ are determined by the specific environment with $C_1+C_2=1$, {\color{black} and $v_{k,m}^{(\mathrm{s})}\triangleq \sin{(\theta_{k,m}^{(\mathrm{s})})} = {z_m}/{d_{k,m}^{(\mathrm{s})}}$ is referred to as the angle indicator.} As such, the {achievable expected rate} $R_{k,m}^{(\mathrm{s})}$ can be approximated by 
\begin{equation}\label{Eq:rateKM}
R_{k,m}^{(\mathrm{s})}\approx\tilde{R}_{k,m}^{(\mathrm{s})} 
\triangleq a_{k,m}^{(\mathrm{s})}\log _{2}\left(1+\frac{p_{k,m}^{(\mathrm{s})} \gamma_{0}f(v_{k,m}^{(\mathrm{s})})}{({z_m^{2}+\|\boldsymbol{q}_{m}-\boldsymbol{u}^{(\rm s)}_{k}\|^{2}})^{\alpha / 2}a_{k,m}^{(\mathrm{s})}}\right).
\end{equation}
Similarly, for the data transmission from UAV $U_m$ to destination node $D_k$, we denote by $a_{m, k}^{(\mathrm{d})} \in[0,1]$ its allocated fraction of bandwidth and $p_{m, k}^{(\mathrm{d})}$ the transmit power of UAV $U_m$. The {achievable expected rate} from $U_m$ to $D_k$ is then approximated by 
\begin{equation}\label{Eq:rateMK}
\begin{aligned}
R_{m,k}^{(\mathrm{d})}\approx \tilde{R}_{m,k}^{(\mathrm{d})}
\triangleq a_{m,k}^{(\mathrm{d})}\log _{2}\left(1+\frac{p_{m,k}^{(\mathrm{d})} \gamma_{0} f(v_{m,k}^{(\mathrm{d})})}{({z_m^{2}+\|\boldsymbol{q}_{m}-\boldsymbol{u}^{(\rm d)}_{k}\|^{2}})^{\alpha / 2}a_{m,k}^{(\mathrm{d})}}\right),
\end{aligned}
\end{equation}
{\color{black} where $v_{m,k}^{(\mathrm{d})}\triangleq \sin{(\theta_{m,k}^{(\mathrm{d})})} = {z_m}/{d_{m,k}^{(\mathrm{d})}}$.}

Next, consider the UAV-UAV communications over LoS channels\footnote{{\color{black}In practical implementation, the UAV-UAV link is established only when the corresponding optimized bandwidth is non-zero.}}. 
Suppose that UAV $U_m$ receives different amounts of data associated with different source nodes and forwards part of them to UAV $U_n$. For each data stream associated with the $k$-th pair ground nodes, we denote by $a_{m,n,k}$ the allocated fraction of bandwidth between UAV $U_m$ and UAV $U_n$, and $p_{m,n,k}$ the transmit power of $U_m$ to $U_n$. {\color{black}Based on the LoS UAV-UAV channel model, the channel power gain can be determined by the link distance without using any channel estimation method.} Then the achievable rate from UAV $U_m$ to $U_n$ for relaying the data of the $k$-th pair ground nodes is given by
\begin{equation}\label{Eq:rateMNK}
\begin{aligned}
R_{m,n,k} &=a_{m,n,k} \log _{2}\left(1+\frac{| h_{m, n}|^{2} p_{m,n,k}}{a_{m,n,k}  N_{0}B\Gamma}\right)=a_{m,n,k} \log _{2}\left(1+\frac{p_{m,n,k} \gamma_{0}}{a_{m,n,k}(\|\boldsymbol{q}_{m}-\boldsymbol{q}_{n}\|^{2}+(z_m-z_n)^2)}\right).
\end{aligned}
\end{equation}

\vspace{-10pt}
\section{Problem Formulation and Proposed Method}\label{PFPSM}
Our objective is to maximize the minimum {achievable expected rate} among all source-destination pairs by jointly optimizing the allocation of bandwidth and transmit power of source nodes and UAVs, as well as UAVs' 3D placement subject to the following constraints. First, let $P_k^{(\rm s)}$ and $P_m$ denote respectively the maximum transmit power of source node $S_k$ and UAV $U_m$. Then the constraints on the transmit power are given by
\begin{align}
&\sum_{m \in \mathcal{M}} p^{(\rm s)}_{k,m} \leq P^{(\rm s)}_{k},~~~~~ \forall k \in \mathcal{K}, \\
&\sum_{k \in \mathcal{K}} \left(p_{m, k}^{(\rm d)}+ \sum_{n \in \mathcal{M}, n\neq m} p_{m,n,k}\right) \leq P_m, ~~~\forall m\in\mathcal{M}.\label{cons:Powm}
\end{align}
Note that in \eqref{cons:Powm}, the total transmit power of each UAV includes that to other UAVs and all destination nodes. Second, as the total bandwidth is orthogonally shared by all source-destination pairs and UAVs, we have the following bandwidth constraint
\begin{equation}
\sum_{k \in \mathcal{K}} \sum_{m \in \mathcal{M}} a_{k,m}^{(\rm s)}\!+\! \sum_{m \in \mathcal{M}} \!\!\left(\sum_{k \in \mathcal{K}} a_{m, k}^{(\rm d)}+\!\!\!  \sum_{n \in \mathcal{M}, n\neq m}\sum_{k \in \mathcal{K}} a_{m,n,k}\right)\! \leq 1.\!\!\!
\end{equation}
Third, the constraints on the UAVs' altitudes are given by
\begin{equation}
    H_{\min}\leq z_m\leq H_{\max}, \forall m\in\mathcal{M}.
\end{equation}
{\color{black}Moreover, we consider the (data) flow conservation constraint for the real-time data relaying, i.e., for each data stream associated with the $k$-th pair of source and destination nodes, a UAV forwards all the data that has been received from the source node and other UAVs.}
{\color{black} As such, for each UAV $U_m$, the flow conservation constraint for the data stream of the $k$-th pair ground nodes can be mathematically expressed as
\begin{align}
& \tilde R_{m, k}^{(\rm d)}+\sum_{n \in \mathcal{M}, n\neq m} R_{m,n,k}=  \tilde R_{k, m}^{(\rm s)}+\sum_{n \in \mathcal{M}, n\neq m} R_{n,m,k}, ~~\forall m\in\mathcal{M}, k\in\mathcal{K}.
\end{align}}
\vspace{-10pt}

For notational convenience, we define $\mathbf{A}\triangleq\{a_{k,m}^{(\rm s)}, a_{m, k}^{(\rm d)}, a_{m,n,k}, \forall k,m,n\}$, $\mathbf{P}\triangleq\{p_{k,m}^{(\rm s)}, p_{m, k}^{(\rm d)}$, $p_{m,n,k},\forall k,m,n\}$, $\mathbf{Q}\triangleq\{\boldsymbol{q}_m, \forall m\}$, $\mathbf{Z}\triangleq\{{z}_m, \forall m\}$, and {\color{black} $\mathbf{V}\triangleq\{v_{k,m}^{(\rm s)}, v_{m, k}^{(\rm d)}, \forall k,m\}$.} {\color{black}Then, the optimization problem for maximizing the minimum {achievable expected rate} among all pairs of ground nodes is formulated as follows.
\allowdisplaybreaks[4]
\begin{subequations}
\begin{align}
\nonumber \textrm{(P1)}\quad\max_{\mathbf{A},\mathbf{P},\mathbf{Q},\mathbf{Z}, \mathbf{V}, \eta}&~~\eta \\
~~~~\textrm{s.t.}
&\sum_{m \in \mathcal{M}} \tilde{R}_{m,k}^{(\rm d)} \geq \eta, \quad \forall k \in \mathcal{K}, \label{cons:ObjOrig}\\
&\sum_{m \in \mathcal{M}} p^{(\rm s)}_{k,m} \leq P^{(\rm s)}_{k},~~~~~ \forall k \in \mathcal{K},\label{cons:PowK}\\
&\sum_{k \in \mathcal{K}} \left(p_{m, k}^{(\rm d)}+ \sum_{n \in \mathcal{M}, n\neq m} p_{m,n,k}\right) \leq P_m, ~~~\forall m\in\mathcal{M},\label{cons:PowM}\\
&\sum_{k \in \mathcal{K}} \sum_{m \in \mathcal{M}} a_{k,m}^{(\rm s)}\!+\! \sum_{m \in \mathcal{M}} \!\!\left(\sum_{k \in \mathcal{K}} a_{m, k}^{(\rm d)}+\!\!\!  \sum_{n \in \mathcal{M}, n\neq m}\sum_{k \in \mathcal{K}} a_{m,n,k}\right)\! \leq 1,\label{cons:Band}\\
&{\color{black}\tilde R_{m, k}^{(\rm d)}\!+\!\!\!\!\!\sum_{n \in \mathcal{M}, n\neq m}\!\!\!\!\!\!\! R_{m,n,k}=\tilde R_{k, m}^{(\rm s)}\!+\!\!\!\!\!\sum_{n \in \mathcal{M}, n\neq m}\!\!\!\!\!\!\! R_{n,m,k} , \forall m\!\in\!\mathcal{M}, k\!\in\!\mathcal{K},\label{cons:InfoOrig}}\\
&v_{k,m}^{(\mathrm{s})}= \frac{z_m}{\sqrt{z_m^{2}+\|\boldsymbol{q}_{m}-\boldsymbol{u}^{(\rm s)}_{k}\|^{2}}},\forall k \in \mathcal{K},m\in\mathcal{M}, \label{eq:vkm}\\
&v_{m,k}^{(\mathrm{d})}= \frac{z_m}{\sqrt{z_m^{2}+\|\boldsymbol{q}_{m}-\boldsymbol{u}^{(\rm d)}_{k}\|^{2}}}, \forall k \in \mathcal{K}, m\in\mathcal{M},\label{eq:vmk}\\
&H_{\min}\leq z_m\leq H_{\max}, \forall m\in\mathcal{M},\label{cons:Alt}
\end{align}
\end{subequations}
where $\tilde R_{k, m}^{(\rm s)}$, $\tilde R_{m, k}^{(\rm d)}$, and $ R_{m,n,k}$ are given in \eqref{Eq:rateKM}, \eqref{Eq:rateMK}, and \eqref{Eq:rateMNK}, respectively.} 

Problem (P1) is generally challenging to solve since the UAVs' 3D placement as well as bandwidth-and-power allocation are coupled in the function of the achievable rate under the UAV-ground Rician fading channel model (see \eqref{Eq:rateKM} and \eqref{Eq:rateMK}), rendering it a highly complicated function that also makes the constraints in \eqref{cons:ObjOrig} and \eqref{cons:InfoOrig} non-convex. Two methods in the literature can be utilized to obtain a suboptimal solution to problem (P1). The first one is the BCD method that iteratively optimizes the bandwidth-and-power allocation as well as UAVs' 3D placement by using convex optimization techniques. 
{\color{black}Note that different from the standard BCD method, the blocks of variables in problem (P1) are coupled in the constraints (see, e.g., \eqref{cons:ObjOrig}, \eqref{cons:InfoOrig}) and thus the feasible set of problem (P1) is not a Cartesian product of the feasible set of individual blocks as in \cite{beck2013convergence}. This may cause the BCD method stuck at a low-quality suboptimal solution and thus suffer substantial rate performance loss, especially when an improper UAVs' placement initialization is adopted.}
In contrast, another method is based on the concept of GS \cite{GSbook}, which progressively learns near-optimal UAVs' 3D placement by stochastically searching in the 3D space for rate maximization.
However, the GS method usually entails an excessively large number of iterations for convergence since it relies on stochastic sampling instead of using the deterministic gradient of the optimization problem as in the BCD method.

\begin{figure}[t]
\centerline{\includegraphics[width=6.8in]{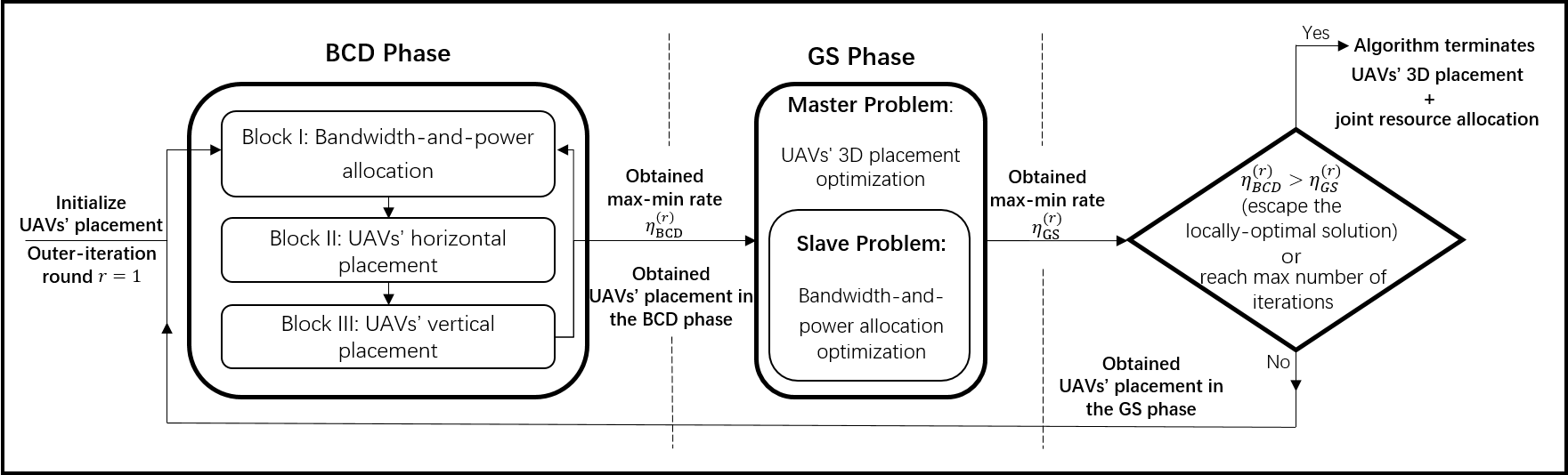}}
\vspace{-18pt}
\caption{Illustration of the proposed IGS-BCD algorithm.}
\label{alg}
\vspace{-18pt}
\end{figure}

Motivated by the above, we propose in this paper a new method to \emph{efficiently} obtain a \emph{high-quality} suboptimal solution to problem (P1), called \emph{IGS-BCD}, by synergizing the advantages of both the deterministic (BCD) and stochastic (GS) methods, namely, \emph{fast convergence} and \emph{superior performance}, respectively. Specifically, the proposed IGS-BCD algorithm alternates between two optimization phases, as illustrated in Fig.~\ref{alg}, which are briefly described as follows and will be elaborated in more details in the next two sections, respectively.

\begin{itemize}
\item[1)] \textbf{BCD phase:} Given initial UAVs' 3D placement (to be specified in Section~\ref{vuc}), this phase aims to quickly find \textit{locally-optimal} UAVs' 3D placement solution by using the BCD method. Specifically, the optimization variables are divided into three sub-problems, namely, the bandwidth-and-power allocation, UAVs' horizontal placement, and UAVs' vertical placement. Then, we iteratively optimize one of the three sub-problems with the other two fixed until all the variables get converged. The obtained solution provides the initial UAVs' 3D placement for the subsequent GS phase.

\item[2)] \textbf{GS phase:} Given UAVs' 3D placement obtained in the BCD phase, the GS phase aims to \emph{evade} the local optimum and further improve the max-min achievable rate by progressively searching the neighboring region of current UAVs' placement in a stochastic manner. Since the bandwidth-and-power allocation is coupled with UAVs' 3D placement in the max-min achievable rate, we reformulate problem (P1) into two sub-problems, corresponding to a slave problem for optimizing the bandwidth-and-power allocation given fixed UAVs' 3D placement and a master problem for optimizing UAVs' 3D placement based on the GS method. Within a prescribed maximum number of iterations, the GS phase will stop and switch to the BCD phase if a better solution than that obtained in the preceding BCD phase is found; otherwise, the IGS-BCD algorithm terminates.

\end{itemize}

\begin{figure}[ht]
\centerline{\includegraphics[width=4in]{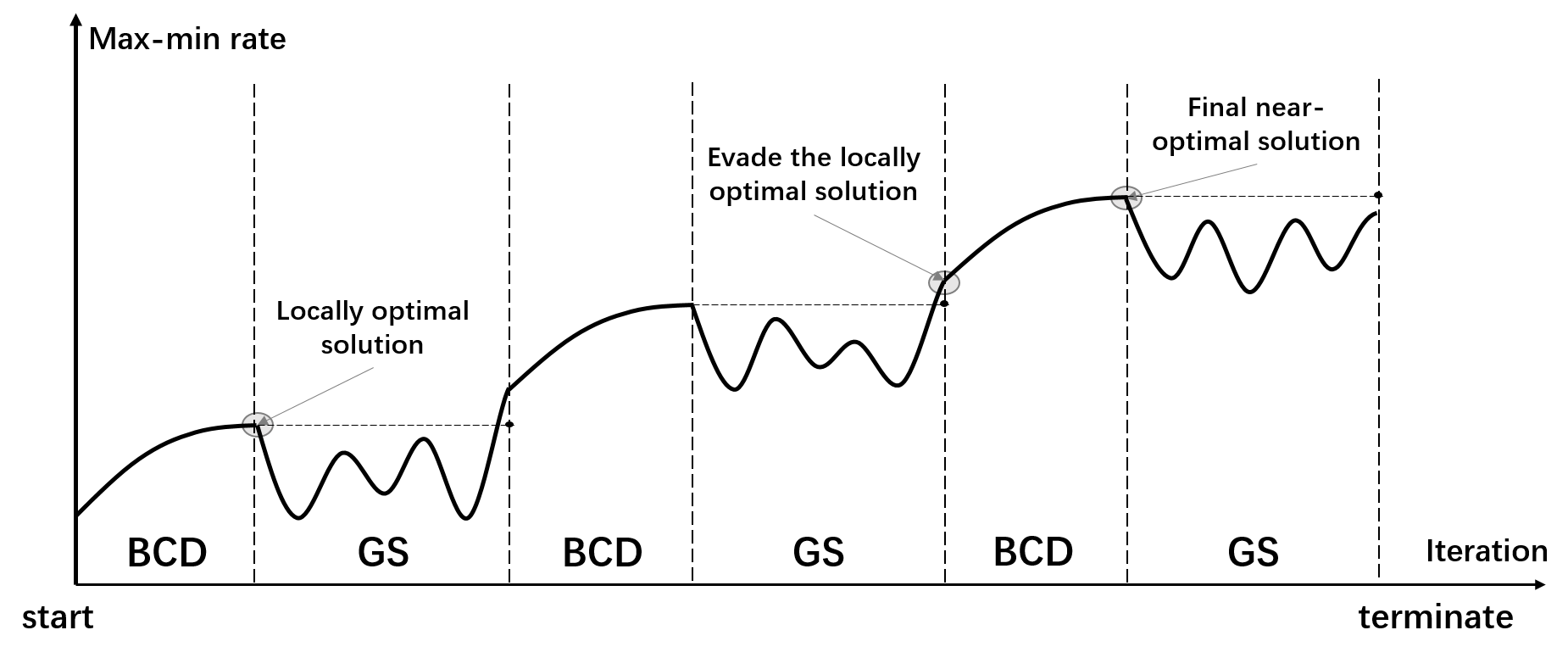}}
\vspace{-18pt}
\caption{Illustration of the max-min rate obtained by the proposed IGS-BCD algorithm over iterations.}
\label{curve}
\vspace{-18pt}
\end{figure}

Moreover, we illustrate in Fig.~\ref{curve} the typical max-min rate obtained by the proposed IGS-BCD algorithm over its iterations. One can observe that, different from the conventional BCD method that may get stuck at a low-quality local optimum, the new method is able to further improve the max-min rate in the subsequent GS phase (albeit that the rate may fluctuate over iterations in each GS phase).

\vspace{-12.5pt}
\section{BCD Optimization Phase}\label{bcd}
\vspace{-4pt}
In this section, we aim to quickly obtain a locally-optimal solution to problem (P1). To this end, we alternately optimize the bandwidth-and-power allocation, UAVs' horizontal placement, as well as their vertical placement by using the BCD method. 

\vspace{-12pt}
\subsection{Bandwidth-and-Power Allocation Optimization}\label{sec:cra}
\vspace{-4pt}

Given any UAVs' 3D placement, the optimization problem (P1) reduces to 
\vspace{-10pt}
\begin{align}
\nonumber \textrm{(P2.a)}\quad \max_{\mathbf{A},\mathbf{P}, \eta} \quad &~~\eta \\
~~~~~~\textrm{s.t.}~~~ \quad 
&\eqref{cons:ObjOrig}-\eqref{cons:InfoOrig}.\nonumber
\end{align}
Note that problem (P2.a) is non-convex due to the non-convex flow conservation constraint in \eqref{cons:InfoOrig}. To tackle this difficulty, we first establish an equivalence between problem (P2.a) and its relaxed problem as follows.
\vspace{-10pt}
{\color{black}
\begin{lemma}\label{lem0}
\emph{Problem (P2.a) can be solved by finding a solution to problem (P2.b) as formulated below that relaxes the equality constraint in \eqref{cons:InfoOrig}.}
\begin{align}
\nonumber \textrm{\emph{(P2.b)}}\quad \max_{\mathbf{A},\mathbf{P}, \eta} \quad &~~\eta \\
~~~~~~\textrm{\emph{s.t.}}~~~ \quad 
&\eqref{cons:ObjOrig}-\eqref{cons:Band},\nonumber\\
&\tilde R_{m, k}^{(\rm d)}\!+\!\!\!\!\!\sum_{n \in \mathcal{M}, n\neq m}\!\!\!\!\!\!\! R_{m,n,k}\leq\tilde R_{k, m}^{(\rm s)}\!+\!\!\!\!\!\sum_{n \in \mathcal{M}, n\neq m}\!\!\!\!\!\!\! R_{n,m,k} , \forall m\!\in\!\mathcal{M}, k\!\in\!\mathcal{K}\label{cons:infor_relax}.
\end{align}
\end{lemma}
\begin{flushleft}
\textit{Proof:} Lemma \ref{lem0} can be proved by contradiction. Specifically, to maximize the minimum achievable rate among all ground nodes in problem (P2.a), the equality in \eqref{cons:infor_relax} for all destination nodes should hold, i.e., $\tilde R_{m, k}^{(\rm d)}+\sum_{n \in \mathcal{M}, n\neq m} R_{m,n,k} = \tilde R_{k, m}^{(\rm s)} +\sum_{n \in \mathcal{M}, n\neq m} R_{m,n,k},$ $\forall m\!\in\!\mathcal{M}$, $k\!\in\!\mathcal{K}$. Otherwise, we can always allocate more power and bandwidth to the UAV-destination links to make the equality hold without decreasing the objective value.
\end{flushleft}}
\vspace{-10pt}
{\color{black}To address the non-convex constraint \eqref{cons:infor_relax}, we then present a useful lemma as below by using the definition of perspective functions \cite{boyd2004convex}.}
\vspace{-10pt}
\begin{lemma}\label{lem1}
\emph{Given $\gamma>0$, the function  $g(x,y)\overset{\triangle}{=}x\log_2 \left(1+\dfrac{\gamma y}{x}\right)$ is jointly concave w.r.t. $x>0$ and $y>0$.}
\end{lemma}
\vspace{-10pt}

Based on Lemma~\ref{lem1}, it can be easily shown that $\tilde R_{m, k}^{(\rm d)} ( R_{m,n,k})$ is concave w.r.t. $p_{m,k}^{(\rm d)}( p_{m,n,k})$ and $a_{m,k}^{(\rm d)}(a_{m,n,k})$. This property allows us to upper-bound $\tilde R_{m, k}^{(\rm d)}$ and $R_{m,n,k}$ as follows by using the SCA technique.

\begin{lemma}\label{lem2}
\emph{Given any UAVs' 3D placement, $\tilde R_{m, k}^{(\rm d)}$ in constraint \eqref{cons:infor_relax} is upper-bounded by}
\begin{equation}\nonumber
\begin{aligned}
&\tilde R_{m, k}^{(\mathrm{d})} \leq \tilde R_{m, k}^{(\mathrm{d}), \mathrm{ub}} \triangleq \hat{R}_{m, k}^{(\mathrm{d})}+\hat{\Psi}_{m, k}^{(\mathrm{d}),\rm{ub}}(a_{m, k}^{(\mathrm{d})}-\hat{a}_{m, k}^{(\mathrm{d})})+\hat{\Phi}_{m, k}^{(\mathrm{d}),\rm{ub}}(p_{m, k}^{(\mathrm{d})}-\hat{p}_{m, k}^{(\mathrm{d})}), \forall m,n\in \mathcal{M},k\in\mathcal{K},\\
\end{aligned}
\end{equation}
\emph{where the equality holds at the point ${a}_{m, k}^{(\mathrm{d})} = \hat{a}_{m, k}^{(\mathrm{d})}$ and ${p}_{m, k}^{(\mathrm{d})}=\hat{p}_{m, k}^{(\mathrm{d})}$. Similarly, we can upper-bound $R_{m,n,k}$ in constraint \eqref{cons:infor_relax} by}
\begin{equation}\nonumber
\begin{aligned}
& R_{m,n,k} \!\leq\! R_{m,n,k}^{\mathrm{ub}}  \!\triangleq\! \hat{R}_{m,n,k}+\hat{\Omega}_{m,n,k}^{\rm{ub}}(a_{m,n,k}-\hat{a}_{m,n,k})\!+\!\hat{\Lambda}_{m,n,k}^{\rm{ub}}(p_{m,n,k}\!-\!\hat{p}_{m,n,k})\!,\! \forall m,n\in \mathcal{M},k\in\mathcal{K},
\end{aligned}
\end{equation}
\emph{where the equality holds at the point ${a}_{m,n,k}=\hat{a}_{m,n,k}$ and ${p}_{m,n,k}=\hat{p}_{m,n,k}$. In the above, $\hat{R}_{m, k}^{(\mathrm{d})}$, $\hat{\Psi}_{m, k}^{(\mathrm{d}),\rm{ub}}$, $\hat{\Phi}_{m, k}^{(\mathrm{d}),\rm{ub}}$, $\hat{R}_{m,n,k}$, $\hat{\Omega}_{m, n}^{(\mathrm{d}),\rm{ub}}$ and $\hat{\Lambda}_{m, n}^{(\mathrm{d}),\rm{ub}}$ are constants that are defined in Appendix A.}
\end{lemma}
\begin{flushleft}
\textit{Proof:} See Appendix A.
\end{flushleft}

\vspace{-8pt}
By using Lemma~\ref{lem2}, problem (P2.b) is approximately reformulated as
\begin{align}
\nonumber \textrm{(P2.c)}\quad \max_{\mathbf{A},\mathbf{P}, \eta} \quad &~~\eta \\
~~~~~~\textrm{s.t.}~~~ \quad 
&\eqref{cons:ObjOrig}-\eqref{cons:Band},\nonumber \\
&\tilde R_{m, k}^{(\rm d), \mathrm{ub}}+\!\!\!\!\!\sum_{n \in \mathcal{M}, n\neq m}\!\!\!\!\!\!\! R_{m,n,k}^{\mathrm{ub}} \leq  \tilde R_{k, m}^{(\rm s)}\!+\!\!\!\!\!\sum_{n \in \mathcal{M}, n\neq m}\!\!\!\!\!\!\! R_{n,m,k}, \forall m\in\mathcal{M}, k\in\mathcal{K}.
\end{align}
{\color{black}It can be easily verified that problem (P2.c) is a convex optimization problem, which can be efficiently solved by using the well-known optimization software, e.g., CVX \cite{cvx}. }

\vspace{-12pt}
\subsection{UAVs' Horizontal Placement Optimization}\label{hpo}
\vspace{-4pt}
Given any feasible resource allocation and UAVs' vertical placement, problem (P1) reduces to the following problem for the UAVs' horizontal placement optimization.
\begin{align}
\nonumber  \textrm{(P3.a)}\quad\max_{\mathbf{Q},\mathbf{V}, \eta} \quad &~~\eta \\
~~~~~~\textrm{s.t.} \quad 
&\eqref{cons:ObjOrig}, \eqref{cons:InfoOrig}-\eqref{eq:vmk}. \nn
\end{align}
{\color{black}Problem (P3.a) is a non-convex optimization problem since the constraints in \eqref{cons:ObjOrig} and \eqref{cons:InfoOrig}--\eqref{eq:vmk} are non-convex}. 
{\color{black}To tackle this difficulty, we first define the following auxiliary variables: $\tilde d_{m,k}^{(\rm d)}={z_{m}^{2}+\left\|\boldsymbol{q}_{m}-\boldsymbol{u}_{k}^{(\mathrm{d})}\right\|^{2}}$, $\tilde d_{m,n,k}=\|\boldsymbol{q}_{m}-\boldsymbol{q}_{n}\|^{2}+(z_m-z_n)^2$, and $\tilde d_{m, k}^{(\rm v)}={1+e^{-(B_{1}+B_{2} v_{m, k}^{(\mathrm{d})})}}$. Then problem (P3.a) is equivalent to
\begin{subequations}
\begin{align}
\nonumber  \textrm{{(P3.b)}}\quad\max_{\mathbf{Q},\mathbf{V},\mathbf{D}, \eta} \quad &~~\eta \\
~~~~~~\textrm{{s.t.}} \quad 
&\sum_{m \in \mathcal{M}} r_{m,k}^{(\rm d)} \geq \eta, \quad \forall k \in \mathcal{K}, \label{cons:ObjRef}\\
& \!r_{m, k}^{(\rm d)}\!+\!\!\!\!\!\sum_{n \in \mathcal{M}, n\neq m}\!\!\!\!\!\!\! r_{m,n,k}= \tilde{R}_{k,m}^{(\rm s)}\!+\!\!\!\!\!\sum_{n \in \mathcal{M}, n\neq m} \!\!\!\!\!\!\!R_{n,m,k},\forall m,n\in\mathcal{M},k\in\!\mathcal{K},\label{eq:InfoRef}\\
&\tilde d_{m,k}^{(\rm d)}= {z_{m}^{2}+\left\|\boldsymbol{q}_{m}-\boldsymbol{u}_{k}^{(\mathrm{d})}\right\|^{2}} , \forall m\in\mathcal{M},k\in\mathcal{K}, \label{eq:dmk}\\
&\tilde d_{m,n,k}=\left\|\boldsymbol{q}_{m}-\boldsymbol{q}_{n}\right\|^{2}+(z_m-z_n)^2, \forall m,n\in\mathcal{M},k\in\mathcal{K}, \label{eq:dmnk}\\
&\tilde d_{m, k}^{(\rm v)} = {1+e^{-(B_{1}+B_{2} v_{m, k}^{(\mathrm{d})})}},\forall m\in\mathcal{M},k\in\mathcal{K},\label{eq:dmkv}\\
&\eqref{eq:vkm},\eqref{eq:vmk},\nonumber
\end{align}
\end{subequations}
where $\mathbf{D} \triangleq \{\tilde d_{m,k}^{(\rm d)},\tilde d_{m,n,k},\tilde d_{m, k}^{(\rm v)},\forall m,n\in \mathcal{M},k\in\mathcal{K}\}$, $r_{m,k}^{(\rm d)}\triangleq a_{m, k}^{(\mathrm{d})} \log _{2}\left(1+\frac{p_{m, k}^{(\mathrm{d})} \gamma_{0} \left({C_{1}+\frac{C_{2}}{d_{m, k}^{(\rm v)}}}\right)}{(\tilde d_{m, k}^{(\mathrm{d})})^{\alpha/2} a_{m, k}^{(\mathrm{d})}}\right)$, and $r_{m,n,k}\triangleq a_{m, n, k} \log _{2}\left(1+\frac{p_{m, n, k} \gamma_{0}}{a_{m, n, k}\tilde d_{m, n, k}}\right)$. Next, we introduce the following important lemma.

\vspace{-3pt}
\begin{lemma}\label{lem3}
\emph{Problem (P3.b) can be solved by finding a solution to problem (P3.c) as formulated below that relaxes the equality constraints in \eqref{eq:vkm}, \eqref{eq:vmk}, and \eqref{eq:InfoRef}--\eqref{eq:dmkv}.}
\vspace{-8pt}
\allowdisplaybreaks[4]
\begin{subequations}
\begin{align}
\nonumber  \textrm{\emph{(P3.c)}}\quad\max_{\mathbf{Q},\mathbf{V},\mathbf{D}, \eta} \quad &~~\eta \\
~~~~~~\textrm{\emph{s.t.}} \quad 
& \!r_{m, k}^{(\rm d)}\!+\!\!\!\!\!\sum_{n \in \mathcal{M}, n\neq m}\!\!\!\!\!\!\! r_{m,n,k}\leq \tilde{R}_{k,m}^{(\rm s)}\!+\!\!\!\!\!\sum_{n \in \mathcal{M}, n\neq m} \!\!\!\!\!\!\!R_{n,m,k},\forall m,n\in\mathcal{M},k\in\!\mathcal{K},\label{cons:InfoRef}\\
&\tilde d_{m,k}^{(\rm d)}\geq {z_{m}^{2}+\left\|\boldsymbol{q}_{m}-\boldsymbol{u}_{k}^{(\mathrm{d})}\right\|^{2}} , \forall m\in\mathcal{M},k\in\mathcal{K}, \label{cons:dmkd}\\
&\tilde d_{m,n,k}\leq\left\|\boldsymbol{q}_{m}-\boldsymbol{q}_{n}\right\|^{2}+(z_m-z_n)^2, \forall m,n\in\mathcal{M},k\in\mathcal{K}, \label{cons:dmnk}\\
&\tilde d_{m, k}^{(\rm v)} \geq {1+e^{-(B_{1}+B_{2} v_{m, k}^{(\mathrm{d})})}},\forall m\in\mathcal{M},k\in\mathcal{K},\label{cons:dmkv}\\
&v_{k, m}^{(\mathrm{s})}\leq \frac{z_m}{\sqrt{z_m^{2}+\|\boldsymbol{q}_{m}-\boldsymbol{u}^{(\rm s)}_{k}\|^{2}}},\forall m\in\mathcal{M},k\in\mathcal{K}, \label{cons:vkm}\\
&v_{m,k}^{(\mathrm{d})}\leq \frac{z_m}{\sqrt{z_m^{2}+\|\boldsymbol{q}_{m}-\boldsymbol{u}^{(\rm d)}_{k}\|^{2}}},\forall m\in\mathcal{M},k\in\mathcal{K}\label{cons:vmk},\\
&\eqref{cons:ObjRef}.\nonumber
\end{align}
\end{subequations}
\end{lemma}
}
\begin{flushleft}
\textit{Proof:} See Appendix B.
\end{flushleft}

\vspace{-8pt}
{\color{black}
Using Lemma 3 in \cite{you20193d}, it can be shown that $r_{m,k}^{(\rm d)}$ is convex w.r.t. $\tilde{d}_{m, k}^{(\mathrm{d})}$ and $\tilde{d}_{m, k}^{(\mathrm{v})}$, $r_{m, n, k}$ is convex w.r.t. $\tilde{d}_{m, n, k}$, $\tilde{R}_{k, m}^{(\mathrm{s})}$ is convex w.r.t. $(1+e^{-\left(B_{1}+B_{2} v_{k, m}^{(s)}\right)})$ and $(z_{m}^{2}+\|\boldsymbol{q}_{m}-\boldsymbol{u}_{k}^{(\mathrm{s})}\|^{2})$, and $R_{n, m, k}$ is convex w.r.t. $\left\|\boldsymbol{q}_{n}-\boldsymbol{q}_{m}\right\|^{2}+(z_n-z_m)^2$. Thus, $r_{m,k}^{(\rm d)}$, $\tilde{R}_{k, m}^{(\rm s)}$ and ${R}_{m,n,k}$ in constraints \eqref{cons:ObjRef} and \eqref{cons:InfoRef} can be approximated by their convex lower-bounds by applying the SCA technique\footnote{{\color{black} There are possibly various upper bounds for $\tilde{R}_{m, k}^{(\mathrm{d})}, \tilde{R}_{k, m}^{(\mathrm{s})}, \text { and } R_{m, n, k}$, (e.g., linear approximation for concave functions), while the one used in the paper is derived from the SCA technique.}}. To address the non-convex constraint \eqref{cons:dmnk}, we define $\ell_{m,n,k}\triangleq\left\|\boldsymbol{q}_{m}-\boldsymbol{q}_{n}\right\|^{2}+(z_m-z_n)^2$ and apply the SCA technique for $\ell_{m,n,k}$.
Moreover, note that constraint \eqref{cons:vkm} can be rewritten as
\begin{align}
\sqrt{z_{m}^{2}+\left\|\mathbf{q}_{m}-\mathbf{u}_{k}^{(\mathrm{s})}\right\|^{2}}+\frac{1}{4} \frac{\left(z_{m}-1\right)^{2}}{v_{k, m}^{(\mathrm{s})}} \leq \frac{1}{4} \frac{\left(z_{m}+1\right)^{2}}{v_{k, m}^{(\mathrm{s})}}\label{con:vkm2},
\end{align} 
where the left-hand side of \eqref{con:vkm2} is a convex function w.r.t. $z_m$ and $v_{k,m}^{\rm{(s)}}$.
The constraint \eqref{cons:vmk} can be rewritten in a similar form as \eqref{con:vkm2}.
In addition, to address the non-convex constraints \eqref{cons:vkm} and \eqref{cons:vmk}, we define $\ell_{k,m}^{(\mathrm{s})} \triangleq \frac{1}{4} \frac{\left(z_{m}+1\right)^{2}}{v_{k, m}^{(\mathrm{s})}}$ and $\ell_{m,k}^{(\mathrm{d})} \triangleq \frac{1}{4} \frac{\left(z_{m}+1\right)^{2}}{v_{m, k}^{(\mathrm{d})}}$, which can be shown to be convex w.r.t. $v_{k, m}^{(\mathrm{s})}$ and $v_{m, k}^{(\mathrm{d})}$, respectively. 
As such, we can lower-bound $\ell_{k,m}^{(\mathrm{s})}$ and $\ell_{m,k}^{(\mathrm{d})}$ with their convex approximations to reformulate constraints \eqref{cons:vkm} and \eqref{cons:vmk} into convex forms.

\vspace{-8pt}
\begin{lemma}\label{lem6} 
\emph{Given any resource allocation and UAVs' vertical placement, $r_{m, k}^{(\mathrm{d})}$, $\tilde R_{k,m}^{(\rm s)}$, ${R}_{m,n,k}$, $\ell_{m, n, k}$, $\ell_{k, m}^{(\mathrm{s})}$, and $\ell_{m, k}^{(\mathrm{d})}$ are lower-bounded by their first-order Taylor expansions as follows.}

\begin{itemize}
    \item $r_{m, k}^{(\mathrm{d})} \geq r_{m, k}^{(\mathrm{d}), \mathrm{lb}}\triangleq \hat{r}_{m, k}^{(\rm {d})}+\hat{\Psi}_{m, k}^{(\mathrm{d}),\rm{lb}}(\tilde{d}_{m, k}^{(\mathrm{d})}- \hat{d}_{m, k}^{(\mathrm{d})})+\hat{\Phi}_{m, k}^{(\mathrm{d}),\rm{lb}}(\tilde{d}_{m, k}^{(\mathrm{v})}-\hat{d}_{m, k}^{(\mathrm{v})}),$ \emph{where} $\!\hat{\Psi}_{m, k}^{(\mathrm{d}),\rm{lb}}=$\newline$-\frac{C_{2 }{\gamma}}{X(X Y^{\frac{\alpha}{2}}+\gamma(C_{1} X+C_{2}))\ln 2}$, $\hat{\Phi}_{m, k}^{(\mathrm{d}),\rm{lb}}=-\frac{\gamma \alpha(C_{1} X+C_{2})}{2 Y(X Y^{\frac{\alpha}{2}}+\gamma(C_{1} X+C_{2}))\ln 2}$ \emph{with $\gamma \triangleq \frac{p_{m,k}^{(\mathrm{d})} \gamma_{0}}{ a_{m,k}^{(\mathrm{d})}}$, $X \triangleq \hat{d}_{m, k}^{(\mathrm{d})}$, $Y \triangleq \hat{d}_{m, k}^{(\mathrm{v})}$, and $\hat{r}_{m, k}^{(\rm {d})}$ is the local value of ${r}_{m, k}^{(\rm {d})}$ at the point $\hat{d}_{m, k}^{(\mathrm{d})}$ and $\hat{d}_{m, k}^{(\mathrm{v})}$. The equality holds at the point $\tilde{d}_{m, k}^{(\mathrm{d})}=\hat{d}_{m, k}^{(\mathrm{d})}$ and $\tilde{d}_{m, k}^{(\mathrm{v})}=\hat{d}_{m, k}^{(\mathrm{v})}$.}
    \newline
    \vspace{-12.5pt}
    \item $\tilde R_{k,m}^{(\rm s)} \geq \tilde R_{k,m}^{(\rm s), \mathrm{lb}}\triangleq \hat{R}_{k,m}^{(\rm s)}+\hat{\Upsilon}_{k,m}^{(\rm s),lb}(e^{-(B_{1}+B_{2} v_{k,m}^{(\mathrm{s})})}- e^{-(B_{1}+B_{2} \hat v_{k,m}^{(\mathrm{s})})})+\hat{\Xi}_{k,m}^{(\rm s),lb}(\|\boldsymbol{q}_{m}-\boldsymbol{u}^{(\rm s)}_{k}\|^{2}-\|\boldsymbol{\hat q}_{m}-\boldsymbol{u}^{(\rm s)}_{k}\|^{2}),$ \emph{where $\hat{\Upsilon}_{k, m}^{(\mathrm{s}), \rm l \mathrm{b}}$ and $\hat{\Xi}_{k, m}^{(\mathrm{s}), \mathrm{lb}}$ can be defined in similar forms as $\hat{\Psi}_{m, k}^{(\mathrm{d}), \mathrm{lb}}$ and $\hat{\Phi}_{m, k}^{(\mathrm{d}), \mathrm{lb}}$, and $\hat{R}_{k,m}^{(\rm s)}$ is the local value of $\tilde{R}_{k,m}^{(\rm s)}$ at the point $\hat v_{k,m}^{(\rm s)}$ and $\boldsymbol{\hat q}_{m}$. The equality holds at the point $v_{k,m}^{(\rm s)}=\hat v_{k,m}^{(\rm s)}$ and $\boldsymbol{q}_{m}=\boldsymbol{\hat q}_{m}$.}
    \newline
    \vspace{-12.5pt}
    \item $R_{m,n,k} \geq R_{m, n, k}^{\mathrm{lb1}}  \triangleq \hat{R}_{m,n,k}+\hat{\Omega}_{m,n,k}^{\rm{lb}}(\|\boldsymbol{q}_{m}-\boldsymbol{q}_{n}\|^{2}-\|\boldsymbol{\hat q}_{m}-\boldsymbol{\hat q}_{n}\|^{2}),$ \emph{where $\hat{\Omega}_{m,n,k}^{\rm{lb}}$ can be defined in a similar form as $\hat{\Psi}_{m, k}^{(\mathrm{d})}$, and $\hat{R}_{m,n,k}$ is the local value of ${R}_{m,n,k}$ at the point $\boldsymbol{\hat q}_{m}$. The equality holds at the point $\boldsymbol{q}_{m}=\boldsymbol{\hat q}_{m}$.}
    \newline
    \vspace{-12.5pt}
    \item $\ell_{m, n, k} \geq \ell_{m, n, k}^{\rm{lb1}}\triangleq \hat{\ell}_{m, n, k} +\hat{\Lambda}_{m, n, k}\left(\left(\boldsymbol{q}_{m}-\boldsymbol{q}_{n}\right)-\left(\boldsymbol{\hat q}_{m}-\boldsymbol{\hat q}_{n}\right)\right)^T,$ \emph{where $\hat{\Lambda}_{m, n, k}=2\left(\boldsymbol{\hat q}_{m}-\boldsymbol{\hat q}_{n}\right)$, and $\hat{\ell}_{k, m}^{(\mathrm{s})}$ is the local value of ${\ell}_{k, m}^{(\mathrm{s})}$ at the point $\boldsymbol{\hat q}_{m}$. The equality holds at the point $\boldsymbol{q}_{m}-\boldsymbol{q}_{n}=\boldsymbol{\hat q}_{m}-\boldsymbol{\hat q}_{n}$.}
    \newline
    \vspace{-12.5pt}
    \item $\ell_{k,m}^{(\rm s)} \geq \ell_{k,m}^{(\rm s),\rm{lb1}}\triangleq \hat{\ell}_{k, m}^{(\mathrm{s})}+\hat{\Lambda}_{k, m}^{(\mathrm{s}), \mathrm{lb1}}(v_{k, m}^{(\mathrm{s})}-\hat v_{k, m}^{(\mathrm{s})}),$ \emph{where $\hat{\Lambda}_{k, m}^{(\mathrm{s}), \mathrm{lb1}}=-\frac{(z_m^2+1)^2}{4(\hat v_{k, m}^{(\mathrm{s})})^2}$, and $\hat{\ell}_{k, m}^{(\mathrm{s})}$ is the local value of ${\ell}_{k, m}^{(\mathrm{s})}$ at the point $\hat v_{k, m}^{(\mathrm{s})}$. The equality holds at the point $ v_{k, m}^{(\mathrm{s})}=\hat v_{k, m}^{(\mathrm{s})}$.}
    \newline
    \vspace{-12.5pt}
    \item $\ell_{m,k}^{(\rm d)} \geq \ell_{m,k}^{(\rm d),\rm{lb1}}\triangleq \hat{\ell}_{m, k}^{(\mathrm{d})}+\hat{\Lambda}_{m, k}^{(\mathrm{d}), \mathrm{lb1}}(v_{m, k}^{(\mathrm{d})}-\hat v_{m, k}^{(\mathrm{d})})$, \emph{where the coefficients $\hat{\Lambda}_{m, k}^{(\mathrm{d}),\rm{lb1}}$ can be defined in a similar form as $\hat{\Lambda}_{k, m}^{(\mathrm{s}), \mathrm{lb1}}$, and $\hat \ell_{m,k}^{(\rm d)}$ is the local value of $\ell_{m,k}^{(\rm d)}$ at the point $\hat v_{m, k}^{(\mathrm{d})}$. The equality holds at the point $ v_{m, k}^{(\mathrm{d})}=\hat v_{m, k}^{(\mathrm{d})}$.}
\end{itemize}
\end{lemma} 
\vspace{-4pt}
The approximation for $r_{m, k}^{(\mathrm{d})}$, $\tilde{R}_{k, m}^{(\mathrm{s})}$, and $R_{m, n, k}$ in Lemma \ref{lem6} can be proved by using the similar method in \cite{you20193d}, with the details omitted for brevity. The approximation for $\ell_{m, n, k}, \ell_{k, m}^{(\mathrm{s})},$ and $\ell_{m, k}^{(\mathrm{d})}$ in Lemma \ref{lem6} is obtained by finding their first-order Taylor expansions. Based on Lemma \ref{lem6}, problem (P3.c) can be transformed into the following approximate form.
\vspace{-10pt}
\begin{subequations}
\begin{align}
\nonumber  \textrm{{(P3.d)}}\quad\max_{\mathbf{Q},\mathbf{V},\mathbf{D}, \eta} \quad &~~\eta \\
~~~~~~\textrm{{s.t.}} \quad 
&\sum_{m \in \mathcal{M}} r_{m, k}^{(\mathrm{d}), \mathrm{lb}} \geq \eta, \quad \forall k \in \mathcal{K},\label{cons:obj_lb}\\
& \!r_{m, k}^{(\rm d)}\!+\!\!\!\!\!\sum_{n \in \mathcal{M}, n\neq m}\!\!\!\!\!\!\! r_{m,n,k}\leq \tilde{R}_{k, m}^{(\mathrm{s}), \mathrm{lb1}}\!+\!\!\!\!\!\sum_{n \in \mathcal{M}, n\neq m} \!\!\!\!\!\!\!R_{m, n, k}^{\mathrm{lb1}},\forall m,n\in\mathcal{M},k\in\!\mathcal{K},\label{cons:flow_lb}\\
&\tilde d_{m,n,k}\leq\ell_{m, n, k}^{\mathrm{lb1}}, \forall m,n\in\mathcal{M},k\in\mathcal{K}, \label{dmnk_lb}\\
&\sqrt{z_{m}^{2}+\left\|\mathbf{q}_{m}-\mathbf{u}_{k}^{(\mathrm{s})}\right\|^{2}}+\frac{1}{4} \frac{\left(z_{m}-1\right)^{2}}{v_{k, m}^{(\mathrm{s})}}\leq \ell_{k, m}^{(\mathrm{s}), \mathrm{lb1}},\forall m\in\mathcal{M},k\in\mathcal{K}, \label{vkm_lb}\\
&\sqrt{z_{m}^{2}+\left\|\mathbf{q}_{m}-\mathbf{u}_{k}^{(\mathrm{d})}\right\|^{2}}+\frac{1}{4} \frac{\left(z_{m}-1\right)^{2}}{v_{m, k}^{(\mathrm{d})}}\leq \ell_{m, k}^{(\mathrm{d}), \mathrm{lb1}},\forall m\in\mathcal{M}, k\in\mathcal{K},\label{vmk_lb}\\
&\eqref{cons:dmkd},\eqref{cons:dmkv}.\nonumber
\end{align}
\end{subequations}
\vspace{-7pt}
Problem (P3.d) is a convex optimization problem and thus can be efficiently solved by using CVX.}

\vspace{-12pt}
\subsection{UAVs' Vertical Placement Optimization}
\vspace{-4pt}
Given any feasible resource allocation and UAVs' horizontal placement, problem (P1) reduces to the UAVs' vertical placement optimization problem as follows.
\begin{align}
\nonumber  \textrm{(P4.a)}\quad\max_{\mathbf{Z},\mathbf{V}, \eta} \quad &~~\eta \\
~~~~~~\textrm{s.t.} \quad 
&\eqref{cons:ObjOrig}, \eqref{cons:InfoOrig}-\eqref{cons:Alt}. \nn
\end{align}

It is observed that problem (P4.a) has a similar form as problem (P3.c). {\color{black}Thus, by following the similar procedures as for solving problem (P3.a), (P4.a) can be transformed into the following approximate form
\vspace{-10pt}
\allowdisplaybreaks[4]
\begin{subequations}
\begin{align}
\nonumber  \textrm{{(P4.b)}}\quad\max_{\mathbf{Z},\mathbf{V},\mathbf{D}, \eta} \quad &~~\eta \\
~~~~~~\textrm{{s.t.}} \quad 
& \!r_{m, k}^{(\rm d)}\!+\!\!\!\!\!\sum_{n \in \mathcal{M}, n\neq m}\!\!\!\!\!\!\! r_{m,n,k}\leq \tilde{R}_{k, m}^{(\mathrm{s}), \mathrm{lb2}}\!+\!\!\!\!\!\sum_{n \in \mathcal{M}, n\neq m} \!\!\!\!\!\!\!R_{m, n, k}^{\mathrm{lb2}},\forall m,n\in\mathcal{M},k\in\!\mathcal{K},\\
&\tilde d_{m,n,k}\leq\ell_{m, n, k}^{\mathrm{lb2}}, \forall m,n\in\mathcal{M},k\in\mathcal{K}, \\
&\sqrt{z_{m}^{2}+\left\|\mathbf{q}_{m}-\mathbf{u}_{k}^{(\mathrm{s})}\right\|^{2}}+\frac{1}{4} \frac{\left(z_{m}-1\right)^{2}}{v_{k, m}^{(\mathrm{s})}}\leq \ell_{k, m}^{(\mathrm{s}), \mathrm{lb2}},\forall m\in\mathcal{M},k\in\mathcal{K}, \\
&\sqrt{z_{m}^{2}+\left\|\mathbf{q}_{m}-\mathbf{u}_{k}^{(\mathrm{d})}\right\|^{2}}+\frac{1}{4} \frac{\left(z_{m}-1\right)^{2}}{v_{m, k}^{(\mathrm{d})}}\leq \ell_{m, k}^{(\mathrm{d}), \mathrm{lb2}},\forall m\in\mathcal{M}, k\in\mathcal{K},\\
&\eqref{cons:dmkd},\eqref{cons:dmkv},\eqref{cons:obj_lb},\nonumber
\end{align}
\end{subequations}
where $\tilde R_{k,m}^{(\rm s), \mathrm{lb2}}\triangleq \hat{R}_{k,m}^{(\rm s)}+\hat{\Upsilon}_{k,m}^{(\rm s),lb}(e^{-(B_{1}+B_{2} v_{k,m}^{(\mathrm{s})})}- e^{-(B_{1}+B_{2} \hat v_{k,m}^{(\mathrm{s})})})+\hat{\Xi}_{k,m}^{(\rm s),lb}(z_m^2-\hat z_m^2)$, $R_{m, n, k}^{\mathrm{lb2}}  \triangleq \hat{R}_{m,n,k}+\hat{\Omega}_{m,n,k}^{\rm{lb}}((z_m-z_n)^2-(\hat z_m-\hat z_n)^2)$, $\ell_{m, n, k}^{\mathrm{lb2}} \triangleq \hat{\ell}_{m, n, k}+2(\hat z_m-\hat z_n)((z_m-z_n)-(\hat z_m-\hat z_n))$, $\ell_{k, m}^{(\mathrm{s}), \mathrm{lb} 2} \triangleq \hat{\ell}_{k, m}^{(\mathrm{s})}+{(z_m-1)(z_m-\hat z_m)}/{2v_{k, m}^{(\mathrm{s})}}-{\left(z_{m}^{2}+1\right)^{2}}\left(v_{k, m}^{(\mathrm{s})}-\hat{v}_{k, m}^{(\mathrm{s})}\right)/{4\left(\hat{v}_{k,m}^{(\mathrm{s})}\right)^{2}}$, and $\ell_{m,k}^{(\mathrm{d}), \mathrm{lb} 2} \triangleq \hat{\ell}_{m,k}^{(\mathrm{d})}+{(z_m-1)(z_m-\hat z_m)}/{2v_{m,k}^{(\mathrm{d})}}-{\left(z_{m}^{2}+1\right)^{2}}\left(v_{m,k}^{(\mathrm{d})}-\hat{v}_{m,k}^{(\mathrm{d})}\right)/{4\left(\hat{v}_{m,k}^{(\mathrm{d})}\right)^{2}}$.}
{\color{black}Problem (P4.b) is a convex optimization problem, which can also be efficiently solved by using CVX. }

\vspace{-12pt}
\subsection{BCD Phase Complexity}\label{bcdcomp}
Based on the results obtained in the preceding three subsections, an iterative BCD algorithm is proposed to obtain a suboptimal solution to problem (P1) by optimizing the resource allocation, UAVs' horizontal placement and their vertical placement one by one with the other two fixed. {\color{black}Moreover, it can be shown that the proposed algorithm in the BCD phase is guaranteed to converge to a stationary solution to problem (P1) \cite{8752072}. Next, we analyze the complexity of the BCD algorithm. 
The convex optimization problems (P2.c), (P3.d), and (P4.b) are not in a standard second-order cone programming (SOCP) form, owing to the logarithm and exponential functions in constraints. To solve problems (P2.c), (P3.d), and (P4.b), a successive approximation method embedded with a primal-dual interior-point method for approximating the logarithm and exponential functions is employed by CVX software.
For example, problem (P3.d) involves $K$ linear inequality constraints of size $2MK+1$ in \eqref{cons:obj_lb}, $MK$ linear inequality constraints of size $5M-1$ in \eqref{cons:flow_lb}, $M^2K$ linear inequality constraints of size $5$ in \eqref{dmnk_lb},  $MK$ second-order cone inequality constraints of size $4$ in \eqref{vkm_lb} and \eqref{vmk_lb}, $MK$ second-order cone inequality constraints of size $3$ in \eqref{cons:dmkd}, and $MK$ second-order cone inequality constraints of size $2$ in \eqref{cons:dmkv}, while the total number of optimization variables is in the order of $M^2K$. Thus, based on the analysis in \cite{6891348}, the worst-case complexity of problem (P3.d) is in the order of $\mathcal{O}\left((M^2K)^{3.5}\log (1 / \epsilon)\right)$, where $\epsilon\geq 0$ represents the prescribed accuracy parameter. In a similar fashion, we can determine the complexities of problems (P2.c) and (P4.b), both of which are in the order $\mathcal{O}\left((M^2K)^{3.5}\log (1 / \epsilon)\right)$ \cite{4518199}.
Then, accounting for the BCD iterations, the overall complexity of each BCD phase is $\mathcal{O}\left(T_{\rm{BCD}}({M^2K})^{3.5}\log(1 / \epsilon)\right)$, where $T_{\rm{BCD}}$ denotes the number of BCD iterations. }

\vspace{-12.5pt}
\section{GS Optimization Phase}\label{gsb}
In this section, we propose a GS-based algorithm to progressively improve the UAVs' 3D placement and resource allocation obtained in the preceding BCD phase that may get stuck at a low-quality local optimum. 
To this end, we decompose problem (P1) into two sub-problems as follows, namely, a slave problem for the resource allocation optimization with given UAVs' 3D placement and a master problem for UAVs' 3D placement optimization.

\subsubsection{Slave problem}
Given any UAVs' 3D placement, the slave problem aims to optimize the bandwidth-and-power allocation for maximizing the minimum achievable rate. This slave problem has the same form as problem (P2.c), and thus can be efficiently solved by using the same method (see Section \ref{sec:cra}). 

\subsubsection{Master problem}
Based on the slave problem (P2.c), the master problem aims to optimize the UAVs' 3D placement for maximizing the minimum achievable rate. Let $\eta(\mathbf{W})$ denote the max-min achievable rate given the UAVs' 3D placement $\mathbf{W}\triangleq\{\mathbf{Q},\mathbf{Z}\}$, where $\eta(\mathbf{W})=\eta^*$ with $\eta^*$ denoting the obtained max-min rate by solving the slave problem (P2.c) with given $\mathbf{W}$. It can be easily shown that the optimal UAVs' 3D placement should be inside the smallest cubic space, denoted by $\boldsymbol{\mathcal{W}}_0$, with its projection on the ground covering all the ground nodes and its altitude confined in \eqref{cons:Alt}. As such, the master problem can be formulated as
\vspace{-3pt}
\begin{align}
\nonumber \textrm{(P5)}~~~~ \max_{\mathbf{W}\in \boldsymbol{\mathcal{W}}_0} \quad &\eta(\mathbf{W}).
\end{align}

Note that the optimal solution to problem (P5) is intractable due to the lack of a closed-form expression for the max-min rate w.r.t. the UAVs' 3D placement, i.e., $\eta(\mathbf{W})$, which can only be computed by solving the slave problem (P2.c) using the iterative algorithm in Section~\ref{sec:cra}. One straightforward approach for solving (P5) is to exhaustively search the UAVs' placement in the cubic space $\boldsymbol{\mathcal{W}}_0$, but this will be computationally costly and even infeasible for a multi-UAV relaying system consisting of a large number of UAVs to cover a large geographical area. To address this issue, we propose to leverage the GS method for progressively finding suboptimal UAVs' placement that is superior to the one obtained in the BCD phase. Specifically, the GS method iteratively updates the UAVs' 3D placement by generating a sequence of samples based on a Markov chain; while each iteration $t$ constitutes $M$ sub-iterations that successively update each UAV's location to its new location according to customized Markov transition probabilities with the locations of other UAVs being fixed. The details of the proposed GS-based algorithm are given as follows. 

First, the cubic space $\boldsymbol{\mathcal{W}}_0$ is equally partitioned into fine-grained small cubic regions with different 3D locations, where the coordinates of the centroids of these regions are denoted by $\mathcal{E}$ as the state space of each UAV's possible locations. Next, we denote $\boldsymbol{w}^{i}_m(t)\in\mathcal{E}$ as the location of UAV $U_m$ in sub-iteration $i$ of the $t$-th iteration, and denote $\boldsymbol{\mathcal W}^{i}(t)$ = $\l\{\boldsymbol{w}_1^{i}(t),\cdots,\boldsymbol{w}^{i}_M(t)\r\}$ as the system state comprising all the $M$ UAVs' current locations. Moreover, let $\boldsymbol{\mathcal W}^{i}_{-m}(t)\!=\!\{\boldsymbol{w}_1^{i}(t),\! \cdots$ $\!,\! \boldsymbol{w}_{m-1}^{i}(t),\boldsymbol{w}_{m+1}^{i}(t),\cdots, \boldsymbol{w}_M^{i}(t)\}$ represent the \textit{partial system state} excluding the location of the $m$-th UAV. The typical realization of the system state $\boldsymbol{\mathcal W}^{i}(t)$ is denoted by $\boldsymbol{W}^{i}(t)\in \mathcal{E}^{M}\overset{\triangle}{=}\underbrace{\mathcal{E}\times \cdots \times \mathcal{E}}_{M}$. Last, the state transition probability from the state in sub-iteration $i-1$ to $i$ in the $t$-th iteration is simply represented by $\Pr[\boldsymbol{\mathcal{{W}}}^{i}(t)|\boldsymbol{\mathcal{{W}}}^{i-1}(t)], i=2, \cdots, M,$ and the transition probability from the state in sub-iteration $M$ of the $(t-1)$-th iteration to sub-iteration $1$ of the $t$-th iteration is denoted  by $\Pr[\boldsymbol{\mathcal{{W}}}^{1}(t)|\boldsymbol{\mathcal{{W}}}^{M}(t-1)]$.

Our objective is to maximize the minimum achievable rate, $\eta({\mathbf{W}})$ of the master problem, by exploring locations around the UAVs' current placement. This is achieved by carefully designing a Markov chain for updating the UAVs' 3D placement as follows.
\begin{itemize}
\item{}{\bf Initialization:} Initialize a UAVs' placement configuration as $\boldsymbol{\mathcal{{W}}}^{1}(t)=\boldsymbol{W}^{1}(t)$ with $t=1$.
\item{}{\bf Sub-iteration:} Successively update the location of each UAV with those of the others being fixed. Specifically, in each sub-iteration $i \in\{2,\cdots, M\}$, the $i$-th UAV is selected for updating its location according to the following state transition probability
\vspace{-3pt}
\begin{align}
&\Pr[\boldsymbol{\mathcal{{W}}}^{i}(t)=\boldsymbol{W}^{i}(t)|\boldsymbol{\mathcal{{W}}}^{i-1}(t)=\boldsymbol{W}^{i-1}(t)]\nn\\
&=\mathbb{I}[\boldsymbol{W}_{-i}^{i}(t)=\boldsymbol{W}_{-i}^{i-1}(t)]\times \frac{e^{\mu \eta(\boldsymbol{W}^{i}(t))}}{\sum_{\boldsymbol{\tilde W}^{i}(t)\in\mathcal{E}^{M}} e^{\mu \eta(\boldsymbol{\tilde W}^{i}(t))}}, ~~~\forall~\!\boldsymbol{W}^{i}(t)\in\mathcal{E}^{M},\label{Eq:TransOri}
\end{align}
where $\mu\ge 0$ is a fixed parameter and $\mathbb{I}[\cdot]$ is an indicator function.
Note that in each sub-iteration $i$, the transition probability for each UAV's placement configuration is jointly determined by its own utility (i.e., the max-min rate) and those of other possible configurations, ${\boldsymbol{\tilde W}^{i}(t)\in\mathcal{E}^{M}}$. The transition probability is non-zero only when the locations of other UAVs except $U_i$ are unchanged. In this way, we only need to adjust one UAV's 3D location in each sub-iteration. In addition, the transition probability of $\Pr[\boldsymbol{\mathcal{{W}}}^{1}(t)|\boldsymbol{\mathcal{{W}}}^{M}(t-1)]$ can be similarly defined and thus are omitted for brevity.
\item{} {\bf Repeat:} Repeat the above sub-iterations multiple times until it evades the local optimum (i.e., the max-min rate obtained in the current sub-iteration of the GS phase is larger than that obtained in the preceding BCD phase\footnote{{\color{black} This guarantees that the converged rate is no smaller than that of the BCD only scheme since the Gibbs sampling phase is employed to find better placement than that obtained by the BCD only scheme to improve the rate.}}) or the maximum number of iterations is reached. In the former case, the GS phase will stop and switch to the BCD phase, while in the latter case, the IGS-BCD algorithm terminates and the final solution is the one obtained in the preceding BCD phase.
\end{itemize}

Note that when the GS phase converges within the prescribed maximum number of iterations, the stationary distribution of the above customized Markov chain is given by
\vspace{-3pt}
\begin{align}
    \lim_{t\to\infty} \Pr\l[\boldsymbol{\mathcal{{W}}}^{M}(t)=\mathbf{W}\r]\overset{\triangle}{=}\pi \l(\mathbf{W}\r)=\frac{e^{\mu \eta(\mathbf{W})}}{\sum_{\mathbf{\tilde W}\in\mathcal{E}^{M}} e^{\mu \eta(\mathbf{\tilde W})}}, ~~\forall~\!\mathbf{W}\in\mathcal{E}^{M}.\label{Eq:StaDis}
\end{align}
This stationary distribution admits a nice property as follows. It is observed that as $\mu\rightarrow \infty$, the stationary probability of the optimal UAVs' placement $\mathbf{W}^*$ for solving problem (P5) is close to $1$, which means that we can obtain the optimal UAVs' 3D placement to problem (P5) after sufficient iterations. In practice, setting too large $\mu$ will incur long time for convergence since it needs more time for environment exploration, while setting too small $\mu$ will incur large optimality gap, whose upper-bound is inversely proportional to $\mu$ \cite{GSbook}. Thus, we need to select a suitable $\mu$ to balance the trade-off between computational time and achievable performance in the UAVs' 3D placement searching. {\color{black}In addition, another critical issue is that devising the customized Markov chain requires calculating the transition probabilities for all possible $\boldsymbol{W}^{i}(t)$ that satisfy $\boldsymbol{W}_{-i}^{i}(t)=\boldsymbol{W}_{-i}^{i-1}(t)$ in \eqref{Eq:TransOri} (corresponding to the case where only the $i$-th UAV's location changes from iteration $i-1$ to iteration $i$). {\color{black}This further necessitates the} computation for the corresponding max-min rates by using the iterative algorithm for resource allocation optimization, which is computationally demanding when the state space becomes large. 

To address this issue, we propose a \emph{refined} GS method as follows that can substantially reduce the computational complexity, and at the same time, achieve high-quality solution. The key idea is to reduce the search space from the entire state space $\mathcal{E}$ to a sub-space that contains two sets of locations. The first set, denoted by $\mathcal{A}_i(t)$, includes the current location of the selected UAV, $\boldsymbol{w}_{i}^{i-1}(t)$, and its neighboring six locations (i.e., the adjacent locations at the upside, downside, front-side, rear-side, left-side, and right-side of $\boldsymbol{w}_{i}^{i-1}(t)$). The second set, denoted by $\mathcal{B}_i(t)$, includes $L\ll |\mathcal{E}|$ random locations in the remaining state space $\mathcal{E}\backslash\mathcal{A}_i(t)$.} Consequently, the transition probability in \eqref{Eq:TransOri} reduces to
\begin{align}
&\Pr[\boldsymbol{\mathcal{{W}}}^{i}(t)=\boldsymbol{W}^{i}(t)|\boldsymbol{\mathcal{{W}}}^{i-1}(t)=\boldsymbol{W}^{i-1}(t)]\nn\\
&=\mathbb{I}[\boldsymbol{W}_{-i}^{i}(t)\!=\!\boldsymbol{W}_{-i}^{i-1}(t)]\!\times\!\mathbb{I}[\boldsymbol{w}_{i}^{i}(t)\in \mathcal{A}_i(t)\!\cup\! \mathcal{B}_i(t) ]\times \frac{e^{\mu \eta(\boldsymbol{W}^{i}(t))}}{\sum_{\boldsymbol{\tilde W}^{i}(t)\in \mathcal{E}^{M}} e^{\mu \eta(\boldsymbol{\tilde W}^{i}(t))}}, \!\forall~\!\boldsymbol{W}^{i}(t)\in\mathcal{E}^{M}.\label{Eq:TransNew}
\end{align}
In practical implementation, for each sub-iteration $i$ of the $t$-th iteration, we only need to calculate the transition probabilities for the states in a reduced accessible space, i.e., $\boldsymbol{W}^{i}(t)\in \mathcal{C}_i(t)\triangleq {\mathcal{E}^{M-1}\times(\mathcal{A}_i(t)\cup \mathcal{B}_i(t))}$. 
It is worth mentioning that the set $\mathcal{A}_i(t)$  is useful for quickly searching a locally better location for the selected UAV and the set $\mathcal{B}_i(t)$ is designed for exploring the entire state space for a potentially better location by random selection. The computational complexity of each GS phase is analyzed as follows. {\color{black}Note that the slave problem (P2.c) in each iteration of the master problem can be solved in parallel with the individual complexity order of $\mathcal{O}\left(({M^2K})^{3.5} \log (1 / \epsilon)\right)$. In each iteration of the master problem, we need to compute the corresponding utility function values of all possible states for each UAV by solving the slave problem and choose the state transition policy according to \eqref{Eq:TransNew}. Thus, the overall complexity order of each GS phase is given by $\mathcal{O}\left(MN({M^2K})^{3.5} \log (1 / \epsilon)T_{\rm{GS}}\right)$, where $N=\vert{\mathcal{C}_i(t)}\vert$ is the cardinality of the reduced accessible space, and $T_{\rm{GS}}$ denotes the maximum number of iterations for the GS phase}.

\vspace{-12.5pt}
\section{Algorithm Initialization and Complexity}
\vspace{-4pt}
In this section, an efficient UAVs' placement initialization scheme is proposed to help accelerate the convergence speed of the proposed IGS-BCD algorithm. Then, we summarize the overall IGS-BCD algorithm for solving problem (P1) and analyze its computational complexity. 

\vspace{-12.5pt}
\subsection{Virtual-UAV Clustering Based Initialization}\label{vuc}
\vspace{-4pt}
To accelerate the convergence speed of the proposed IGS-BCD algorithm, we propose a new UAVs' placement initialization scheme, called \emph{virtual-UAV clustering} (VUC), accounting for the spatial distribution of source-destination nodes as well as the asymmetry in the practical transmit power of ground nodes and UAVs. The main procedures are presented as follows.

\begin{itemize}
\item[1)] \textbf{Virtual-UAV placement}: First, we assume that each pair of source and destination nodes is assigned with one virtual UAV at an initial altitude denoted by \!$H_0\in [H_{\rm{min}},H_{\rm{max}}]$ for assisting data relaying. We aim to determine the horizontal placement of \!$K$\! virtual UAVs, which are denoted by $\left\{\widetilde{U}_{\tilde{k}}, \tilde{k} \in \mathcal{K}\right\}$, under the constraints on the bandwidth-and-power allocation. 
For ease of design, we assume equal bandwidth-and-power allocation, i.e., the transmit power of each virtual UAV is $(\sum_{m=1}^{M}\! P_m)/K$, and the bandwidth of each UAV-UAV and UAV-ground link is $B/(2K)$. 
The transmit power at each source node is set as its maximum value. Then, for each pair of nodes, we optimize the virtual-UAVs' locations to maximize the individual source-destination achievable rate by solving the following problem.
\vspace{-18pt}
\begin{align}
\nonumber \textrm{(P6)}~~~~ \max_{\mathbf{\tilde{\mathbf Q}}} \quad &R_{\tilde{k},k}^{(\rm d)}\\
~~~~~~\textrm{s.t.} \quad 
&R_{\tilde k,k}^{(\rm d)}\leq R_{k,\tilde{k}}^{(\rm s)},\forall k, \tilde{k}\in\mathcal{K}, \label{IniCons}
\end{align}
where $\tilde{\mathbf Q}$ is the horizontal locations of virtual UAVs, $R_{k,\tilde{k}}^{(\rm s)}$ is the maximum achievable rate from source node $S_k$ to virtual UAV $\widetilde{U}_{\tilde{k}}$, and $R_{\tilde k,k}^{(\rm d)}$ is the maximum achievable rate from virtual UAV $\widetilde{U}_{\tilde{k}}$ to destination node $D_k$. It can be easily shown that for the optimal solution to (P6), the equality in constraints \eqref{IniCons} should hold. This corresponds to searching for a point on the straight line connecting the two ground nodes that achieves  $R_{\tilde k,k}^{(\rm d)}= R_{k,\tilde k}^{(\rm s)}$.

\item[2)] \textbf{Virtual-UAV clustering}: Second, the initial UAVs' placement is determined as the centroids of $M$ clusters of the $K$ virtual UAVs by using the K-means clustering \cite{1017616}.
\end{itemize}

Note that directly deploying the UAVs at the centroids of $M$ clusters of the $2K$ ground nodes may be inappropriate in our case. For instance, when the source and destination nodes form distant clusters, this method will result in no UAV relays being deployed between these clusters and thus limit the max-min rate. This issue is addressed by our proposed initialization scheme with a virtual-UAV placement followed by a virtual-UAV clustering.

\vspace{-18pt}
\subsection{Overall Algorithm and Complexity Analysis}
\vspace{-4pt}
The proposed IGS-BCD algorithm starts with the initial UAVs' placement as in Section~\ref{vuc} and then alternates between the BCD and GS phases until we cannot find a better solution within a prescribed maximum number of iterations in the GS phase. It is worth noting that when the BCD phase switches to the GS phase, the initial UAVs' placement locations are set as the discrete centroids of the cubic fine-grained regions that are closest to the continuous ones obtained in the preceding BCD phase.

Next, we discuss the complexity of the overall IGS-BCD algorithm. Let \!$T_{\rm{out}}$\! denote the total number of outer iterations of the BCD and GS phases. {\color{black}In each outer iteration, the BCD and GS phases have individual complexity orders of $\mathcal{O}(T_{\rm{BCD}}(M^2K)^{3.5} \log (1 / \epsilon))$ (see Section~\ref{bcdcomp}) and $\mathcal{O}(MN(M^2K)^{3.5} \log (1 / \epsilon)T_{\rm{GS}})$ (see Section~\ref{gsb}), respectively. Thus, the total computational complexity order of our proposed IGS-BCD algorithm is dominated by the GS phase, i.e., $\mathcal{O}(T_{\rm{out}}MN(M^2K)^{3.5} \log (1 / \epsilon)T_{\rm{GS}})$\footnote{{\color{black} Although it is intractable to analyze the convergence rate of the proposed scheme, it is expected that the proposed scheme converges faster than the GS only scheme, since it uses convex optimization techniques to find the high-quality UAVs' placement more efficiently instead of randomly searching for better solutions as in the GS only scheme, which is also corroborated by our simulation results.}}.}

\vspace{-12.5pt}
\section{Simulation Results}\label{sr}
\begin{figure}[ht]
\centerline{\includegraphics[width=2.8in]{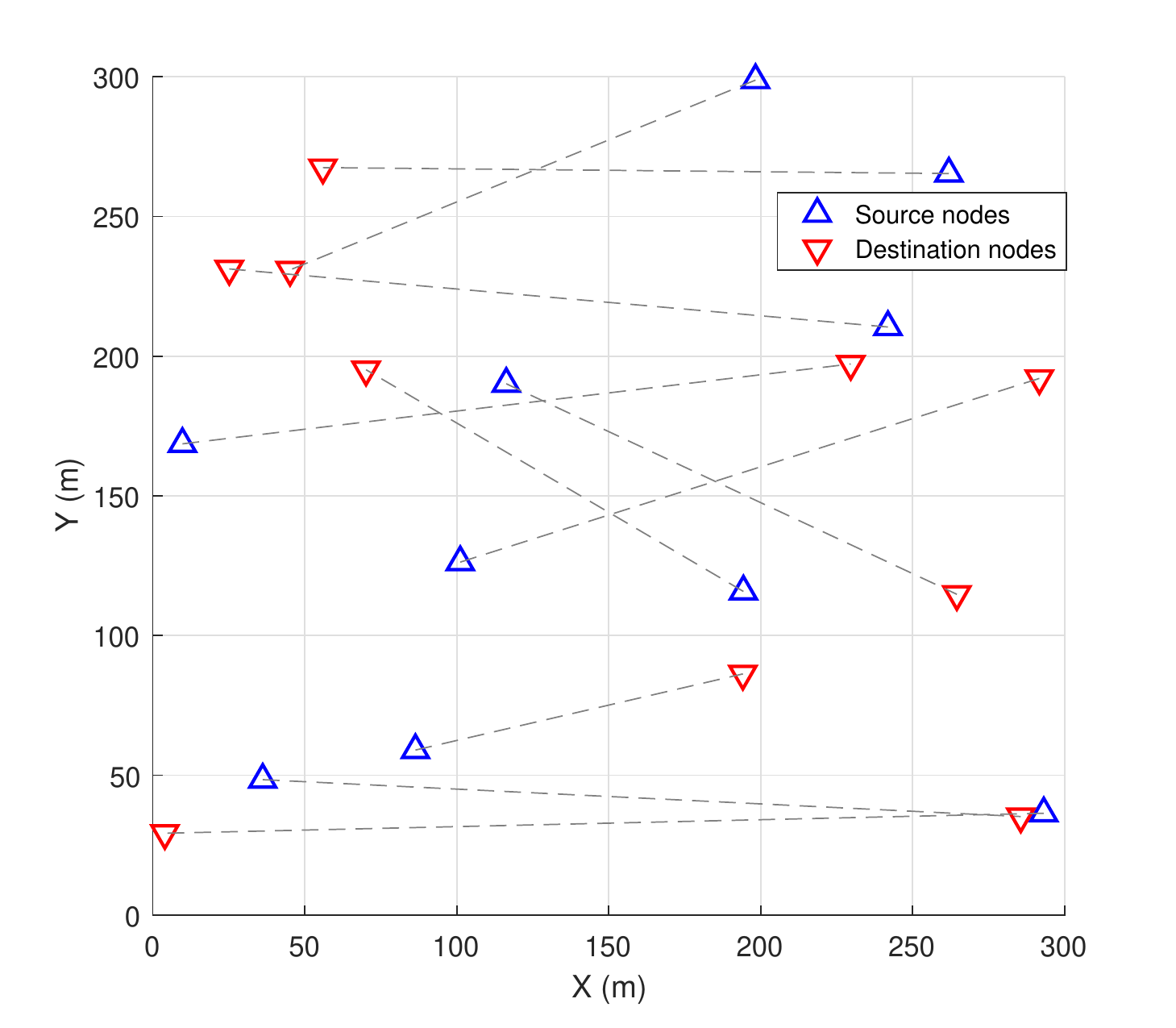}}
\vspace{-12.5pt}
\caption{Locations of ground source-destination pairs for simulation.}
\label{gdloc}
\vspace{-18pt}
\end{figure}

\vspace{-4pt}
Simulation results are presented in this section to verify the effectiveness of the proposed algorithm. For ease of illustration, without otherwise specified, we consider a random realization of 10 pairs of source and destination nodes distributed in a $300\times 300~ \text{m}^2$ square area, as shown in Fig.~\ref{gdloc}. Three UAVs are deployed to assist data relaying for ground nodes in a $300\times300\times150~ \text{m}^3$ 3D space that is equally partitioned into equal-size cubes of $5\times5\times5~ \text{m}^3$ in the GS phase. We assume that all the UAVs have the same maximum transmit power of $P_m=2$ W, $\forall m$, and all the source nodes have the same maximum transmit power of $P_k^{\rm(s)}=15$ dBm, $\forall k$. The flying altitude for all UAVs is limited in the range of $[30,150]$ m. The total system bandwidth is $B=10$ MHz, the received noise power density is $N_0=-169$ dBm/Hz, and the SNR gap $\Gamma=8.2$ dB. The parameters for the Rician fading channel model are set as $B_1 = -4.3224, B_2 = 6.0750, C_1=0$, and $C_2=1$ \cite{you20193d}. In the UAVs' placement initialization, all UAVs' initial altitudes are set as $50$ m. For the GS phase, we set the number of searched random locations $L = 3$. Other parameters are set as $\alpha=2.5$, $\beta_0=-30$ dB, $\mu=30$, and $\epsilon=0.001$.
\vspace{-12.5pt}
\vspace{-12.5pt}
\subsection{Algorithm Performance}
\vspace{-4pt}
\subsubsection{Performance of the proposed initialization scheme}
We first compare our proposed VUC-based initialization with two benchmark initializations: 1) Random initialization: The UAVs are randomly deployed in the equally-partitioned cubes; 2) Ground node clustering (GNC) based initialization: The UAVs are placed at the three cluster centroids of the ground nodes. 

\begin{figure}[t] \centering  
{\color{black}\subfigure[{\color{black}Instantaneous max-min rate over iterations.}] {
\label{ealg7}
\includegraphics[width=2.8in]{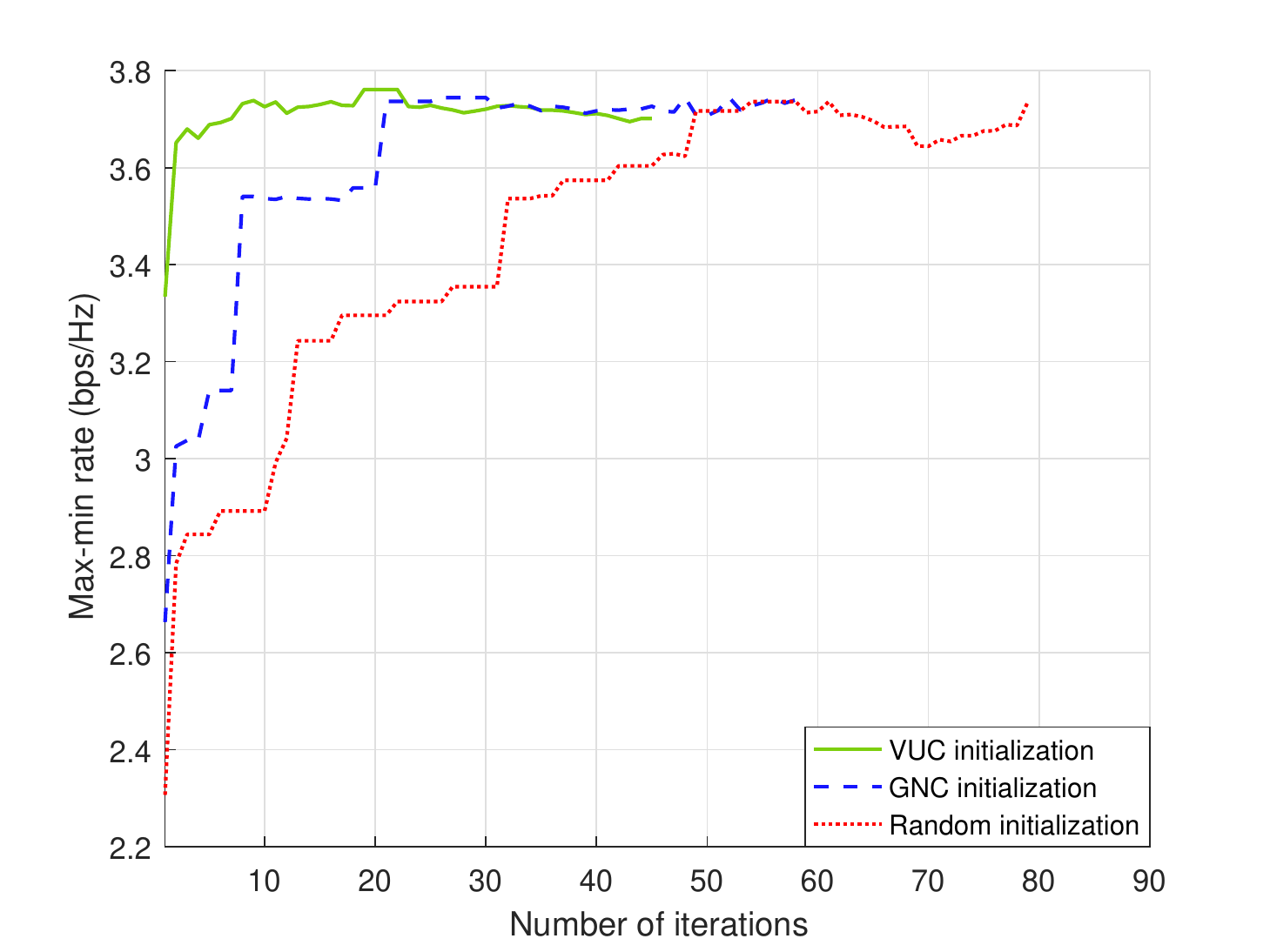}  
\vspace{-18pt}
}}     
{\color{black}\subfigure[{\color{black}Accumulatively best max-min rate over iterations.}] { 
\label{ealg8}
\includegraphics[width=2.8in]{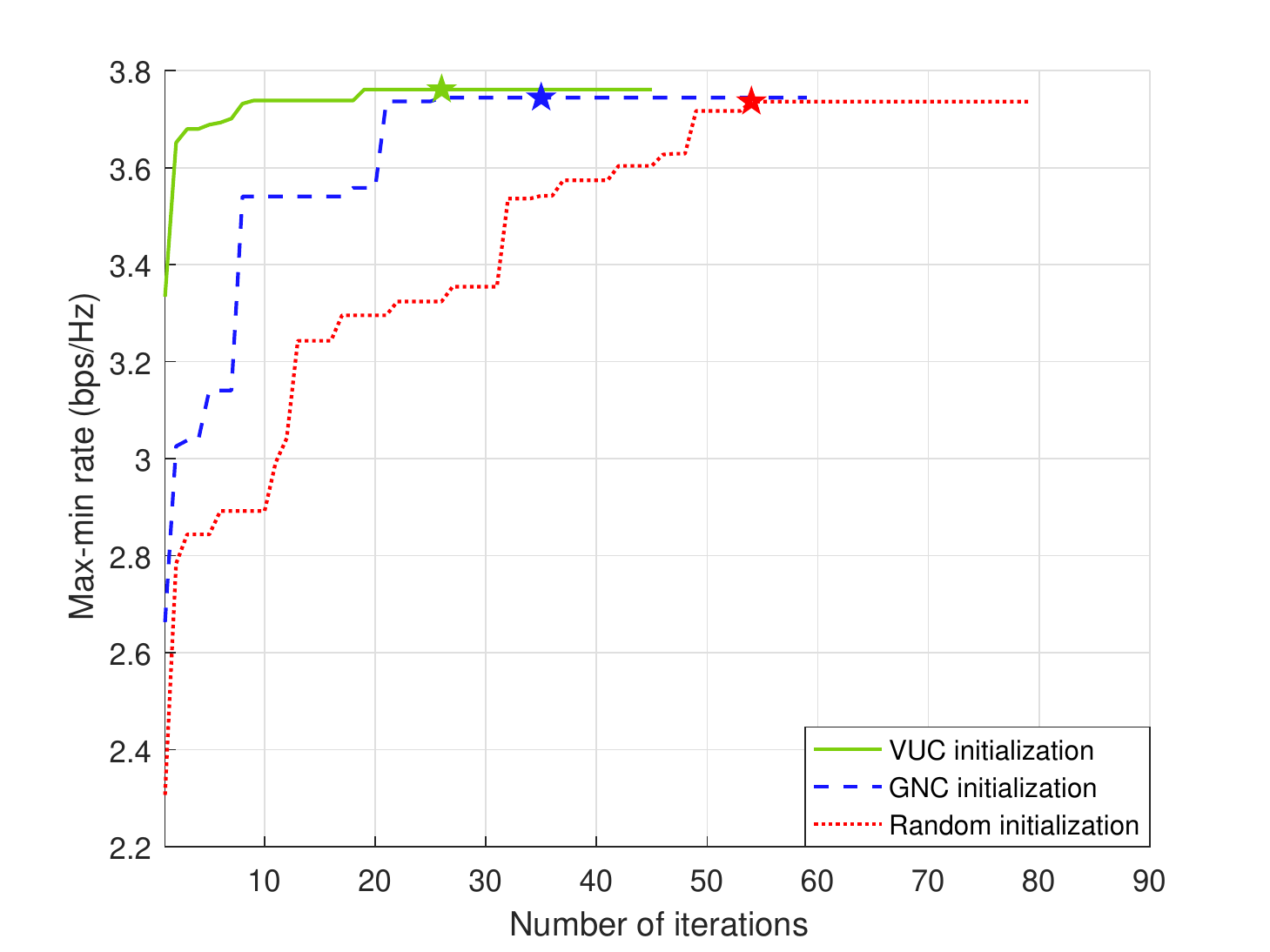}   
\vspace{-18pt}
}}

{\color{black}\caption{Convergence performance comparison of the IGS-BCD algorithm with different initializations.}} 
\label{res3}     
\vspace{-18pt}
\end{figure}

\begin{figure}[t]
\centerline{\includegraphics[width=2.8in]{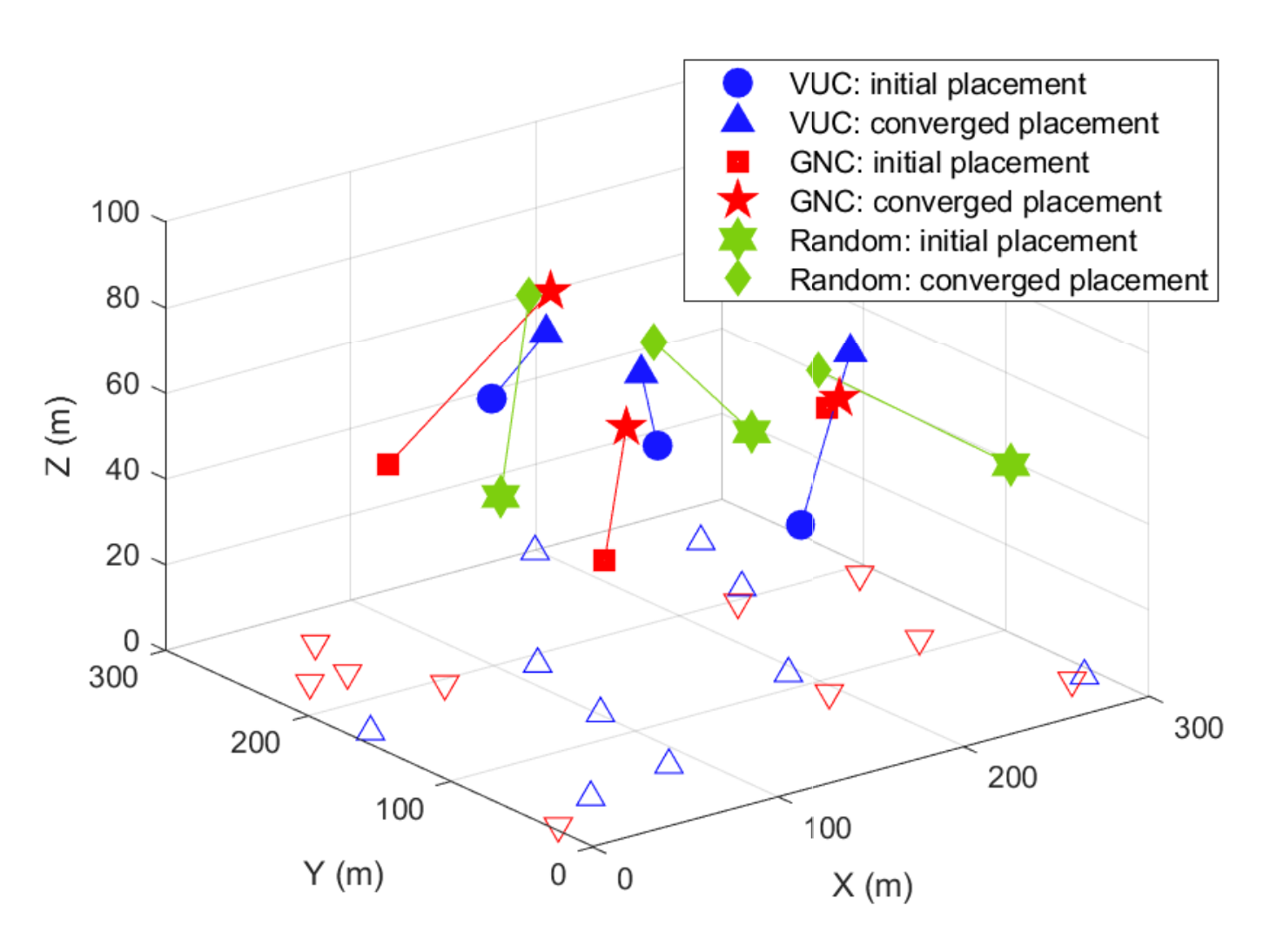}}
\vspace{-18pt}
{\color{black}\caption{Converged UAVs' 3D placement by the IGS-BCD algorithm with different initializations.}}
\label{res4}
\vspace{-18pt}
\end{figure}

In Figs.~\ref{ealg7} and \ref{ealg8}, we compare the instantaneous max-min rate over the (outer) iterations as well as the accumulatively best max-min rate over the so-far conducted iterations, respectively. First, it is observed that the IGS-BCD algorithm with our proposed VUC-based initialization converges faster than that with benchmark initializations. In addition, under different UAVs' placement initializations, our proposed IGS-BCD algorithm is observed to achieve similar converged max-min rates, which demonstrates its robustness against different UAVs' placement initialization schemes.

Fig.~\ref{res4} shows the converged UAVs' 3D placement by the proposed IGS-BCD algorithm with different initializations. It is observed that the converged UAVs' locations are similar, regardless of their initial locations, which is in accordance with the similar converged max-min rates of different initializations observed in Fig.~\ref{res3}.

\subsubsection{Performance of the proposed IGS-BCD algorithm}
Next, we demonstrate the effectiveness of the proposed IGS-BCD algorithm with VUC-based initialization as compared to the following benchmark schemes: 1) Random placement with selection: Randomly generate 300 sets of UAVs' 3D placement in the cubic region of interest with optimized communication resource allocation and select the one that achieves the largest max-min rate; 2) BCD only with VUC initialization: Apply the BCD method only for solving problem (P1) with VUC-based initialization of UAVs' placement; 3) BCD only with multiple initializations: Apply the BCD method only for solving problem (P1) with 100 initializations of UAVs' placement and select the converged placement that achieves the largest max-min rate. In particular, the first UAVs' initial placement is obtained by the VUC-based initialization for fair comparison; while the subsequent initializations are randomly selected from the neighboring region of the converged placement with the first VUC-based initialization; 4) GS only: Search UAVs' 3D placement by using the GS method only with VUC-based initialization of UAVs' placement.

\begin{figure}[t] \centering    
{\color{black}\subfigure[{\color{black}Accumulatively best max-min rate versus computational time.}] {
\label{res1}
\includegraphics[width=2.9in]{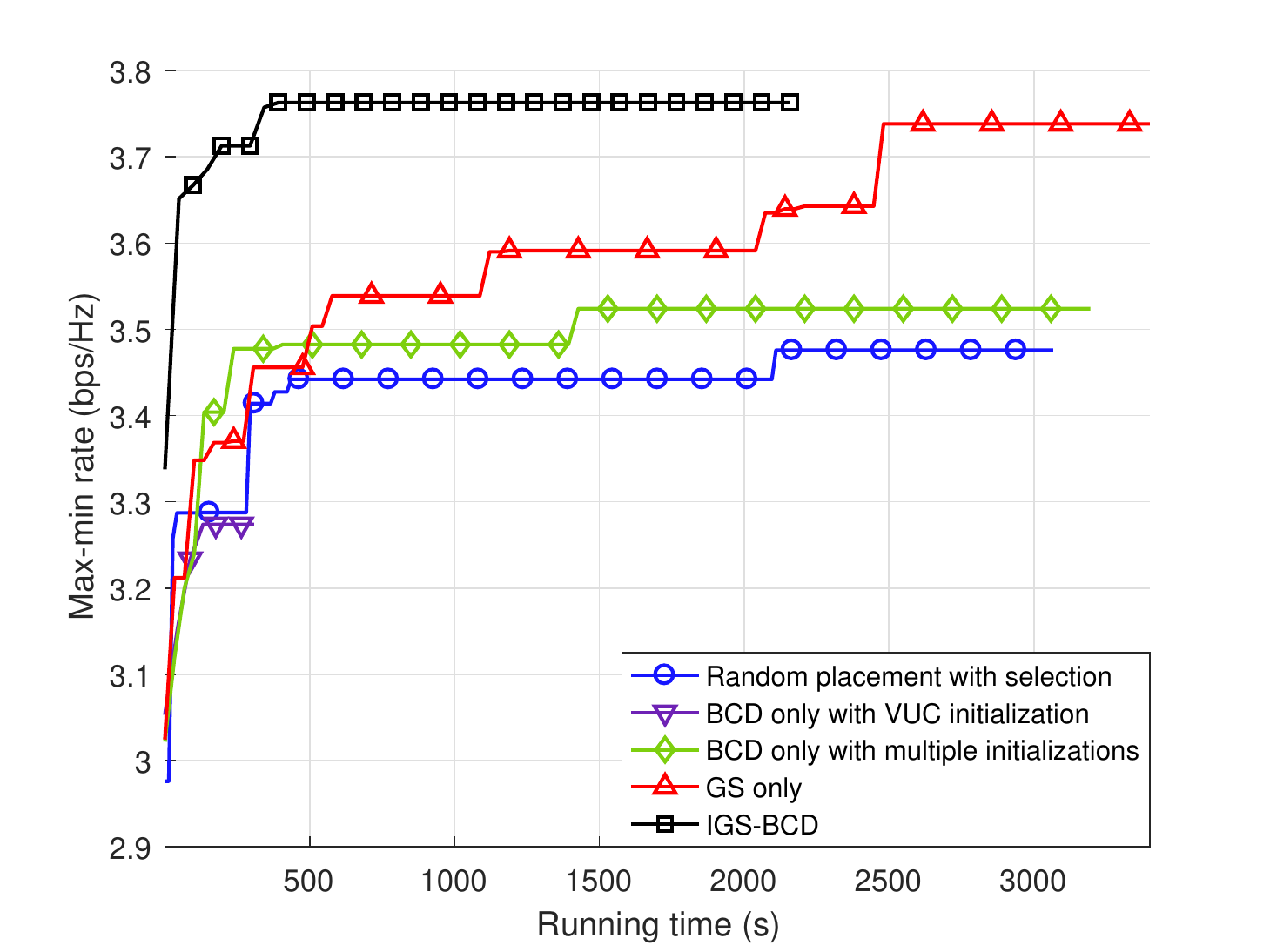}  
\vspace{-18pt}
}}     
{\color{black}\subfigure[{\color{black}Converged UAVs’ 3D placement.}] {\label{res2}
\includegraphics[width=2.9in]{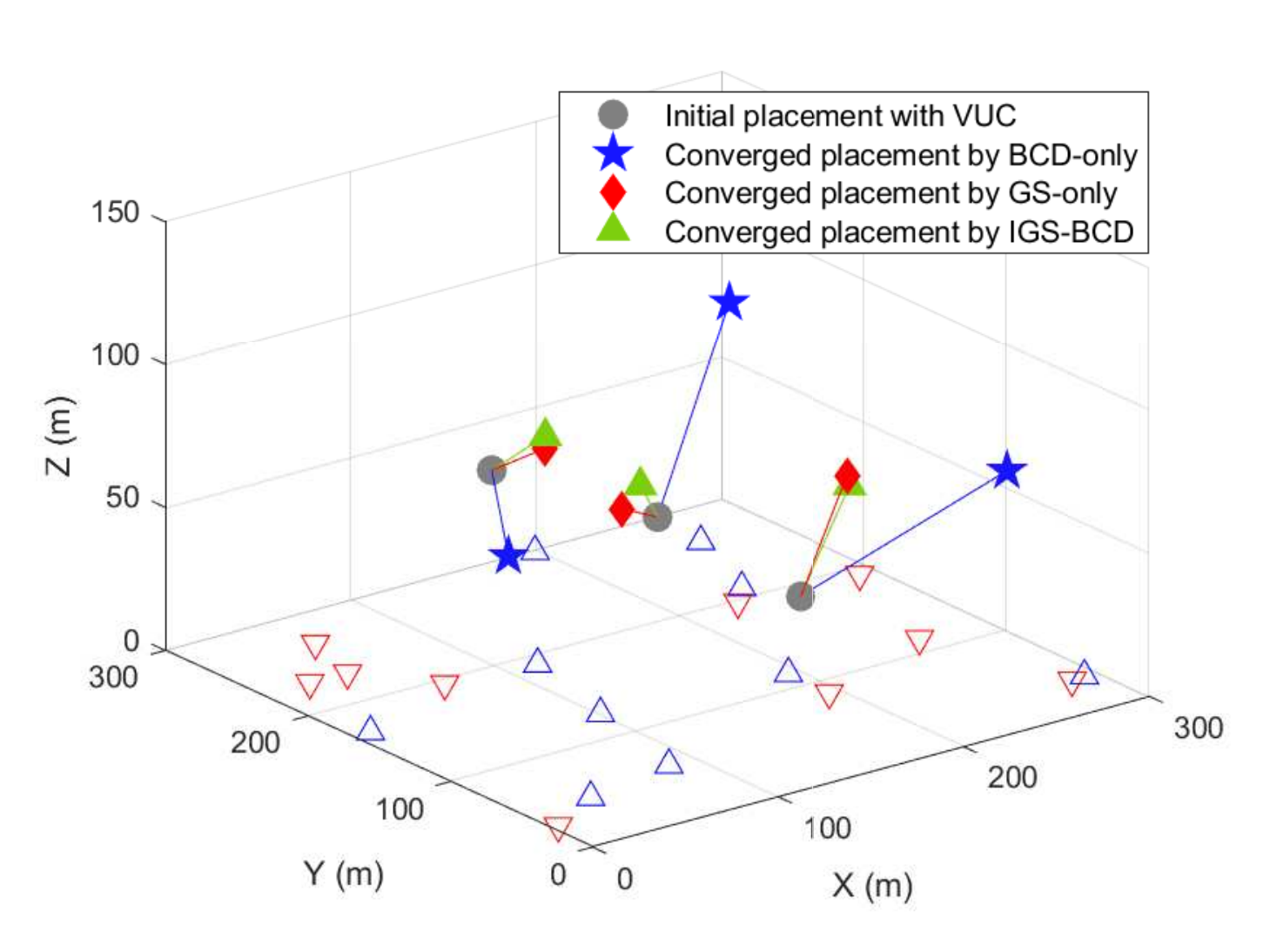}     
\vspace{-18pt}
}}    

{\color{black}\caption{Comparison of the rate performance and UAVs' 3D placement by different schemes.}}
\vspace{-22pt}
\end{figure}

Fig.~\ref{res1} compares the converged max-min rate and required computational time{\footnote{\color{black}The actual running time of the proposed algorithm implemented in a real-time system is in general shorter than that is solved using CVX since CVX involves additional overhead to reformulate problems into standard forms.}} of different schemes using Matlab 2019a on a computer with Intel i5 3.4 GHz CPU and 8-GB memory. Several important observations are made as follows. First, our proposed IGS-BCD algorithm and the GS-only scheme achieve similar max-min rates, which significantly outperform other benchmark schemes. This is expected since they both avoid getting stuck at local optimum via searching unexploited UAVs' locations. Second, it is observed that the proposed IGS-BCD algorithm converges significantly faster than the GS-only scheme, due to the use of BCD to execute the local-optimum computation more efficiently.
Third, it is observed that the BCD-only scheme with multiple initializations can only improve the max-min rate marginally after the BCD optimization based on the initial VUC-based initialization because the employed local research is inefficient as compared to the GS-based search in our proposed IGS-BCD algorithm.

Fig.~\ref{res2} shows the converged UAVs' 3D placement of different schemes, all using the VUC-based initialization. It is observed that given the same initialization, the converged UAVs' 3D placement of the proposed IGS-BCD algorithm is close to that of the GS-only scheme, which is expected as they achieve similar max-min rates as shown in Fig.~\ref{res1}, whereas that of the BCD-only scheme is substantially different (thus resulting in suboptimal max-min rate performance as shown in Fig.~\ref{res1}).
{\color{black} In addition, it is observed that the altitudes of the converged UAVs' placement are elevated as compared to those of initial placement. 
This is because at lower altitude, the UAV-ground elevation angle is relatively small initially, which results in more multi-path fading under the considered elevation-angle dependent Rician fading channel model. However, our proposed scheme properly increases the UAVs' altitudes to enlarge the elevation angles with slightly higher path loss so as to better balance the angle-versus-distance tradeoff.}

\vspace{-18pt}
\subsection{Effects of System Parameters}
\vspace{-4pt}
Next, we evaluate the effects of some key system parameters on the rate performance and UAVs' placement.

\subsubsection{Effects of the numbers of source-destination pairs or UAVs}

\begin{figure}[t] \centering    
{\color{black}\subfigure[Max-min rate versus the number of source-destination pairs.] {
\label{res51}
\includegraphics[width=3.1in]{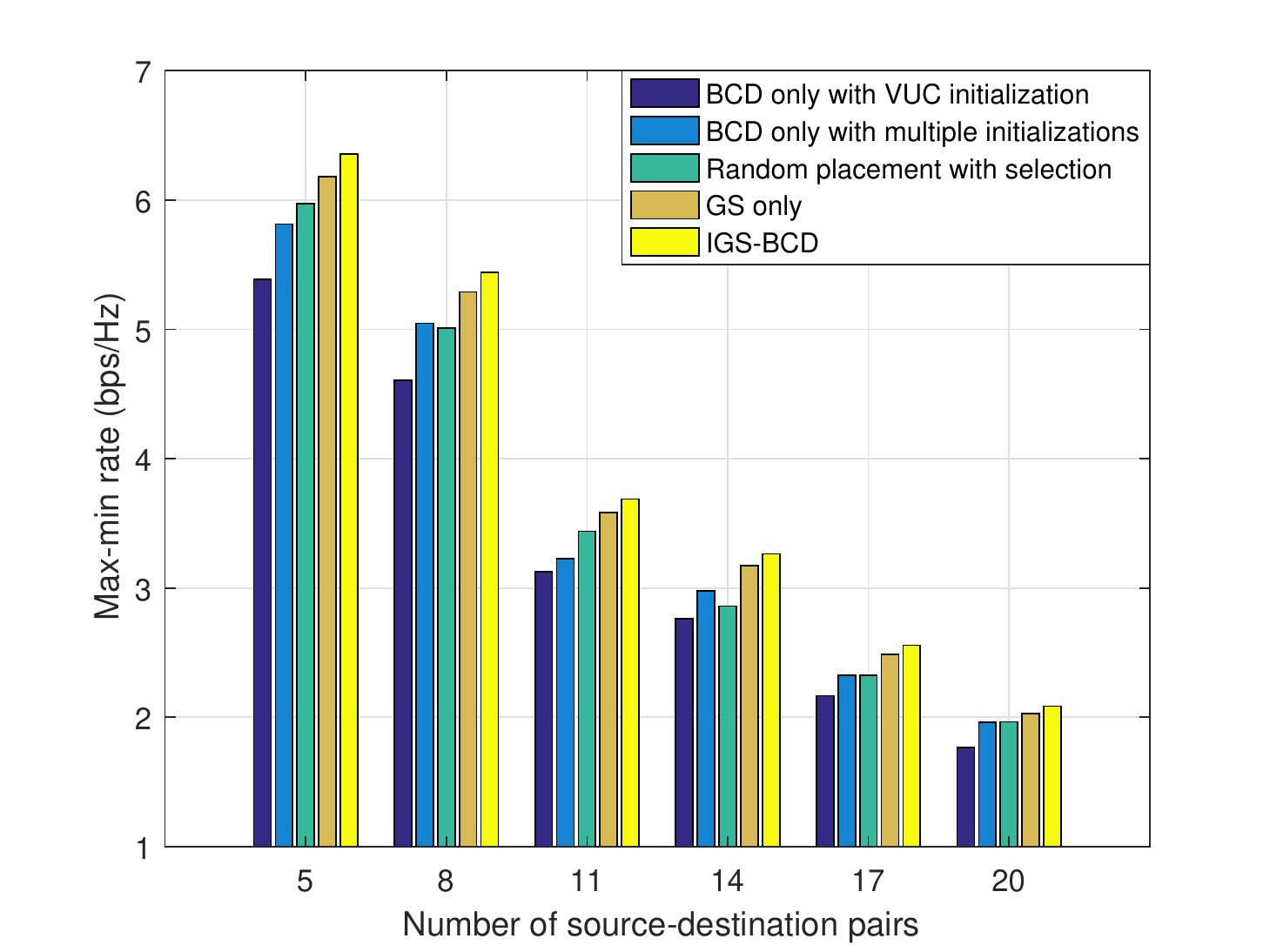}  
}}     
{\color{black}\subfigure[Max-min\! rate\! versus\! the\! number\! of\! UAVs.\!] { 
\label{res52}
\includegraphics[width=3.1in]{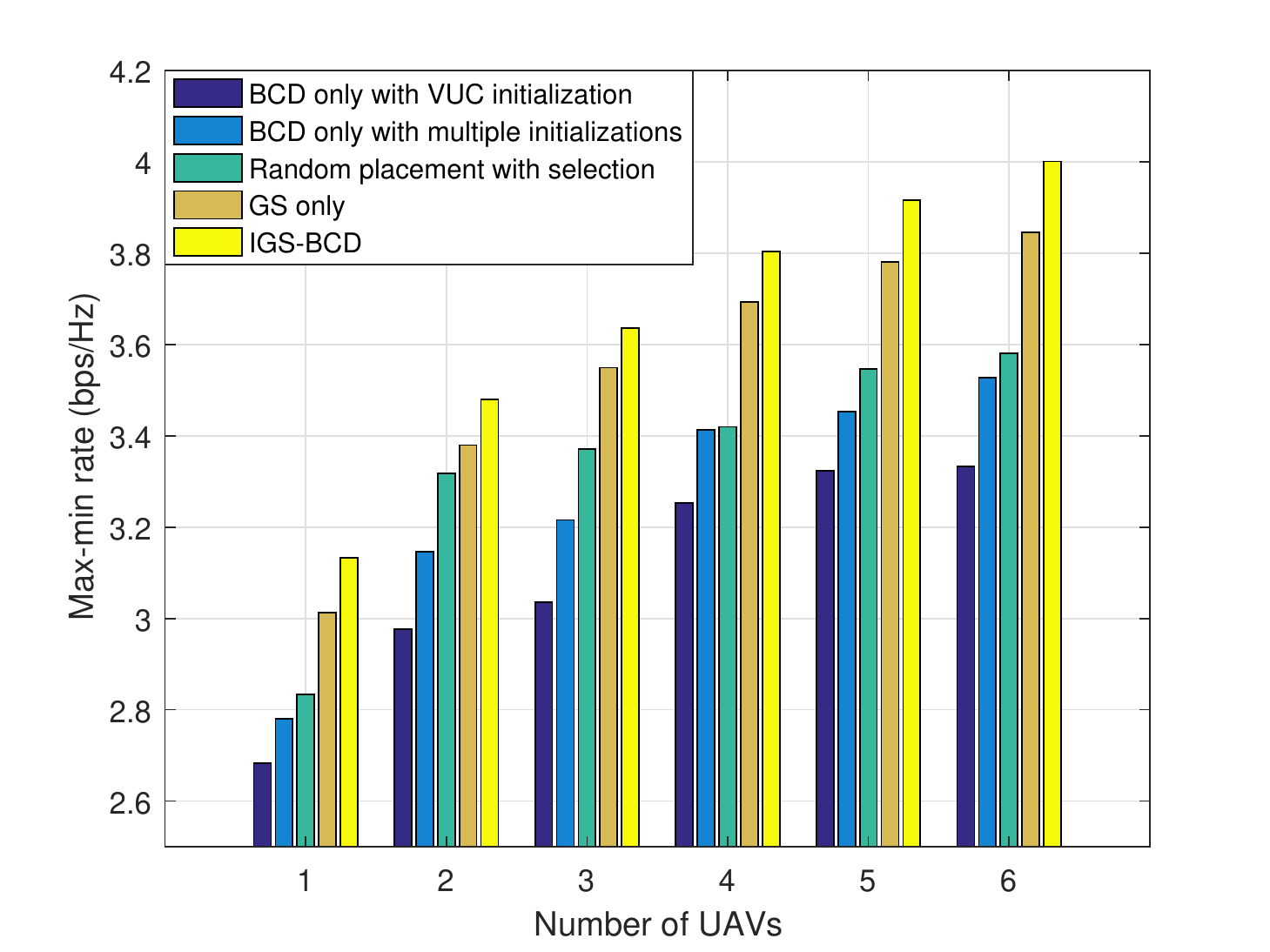}     
}}    

{\color{black}\caption{Effects of the number of source-destination pairs and UAVs.}  }
\label{res5}     
\vspace{-22pt}
\end{figure}

In Fig.~\ref{res51}, we plot the converged max-min rates by different schemes versus the number of source-destination pairs, $K$. It is observed that given a fixed number of UAVs, the max-min rate monotonically decreases with the increasing number of ground nodes, which is expected since the total bandwidth and the transmit power of UAVs for data relaying is limited. Our proposed IGS-BCD algorithm is observed to achieve a larger max-min rate over the benchmark schemes for all the values of $K$. In Fig.~\ref{res52}, we compare the achieved max-min rates by different schemes versus the number of UAVs, $M$. Similarly, it is observed that all the schemes achieve larger max-min rate with the increasing number of UAVs, and the proposed IGS-BCD algorithm outperforms the benchmark schemes for all the values of $M$.

\subsubsection{Effects of ground nodes' spatial distribution}

\begin{figure}[t] \centering    
\subfigure[Inter-cluster communication.] {
\label{bc}
\includegraphics[width=3.1in]{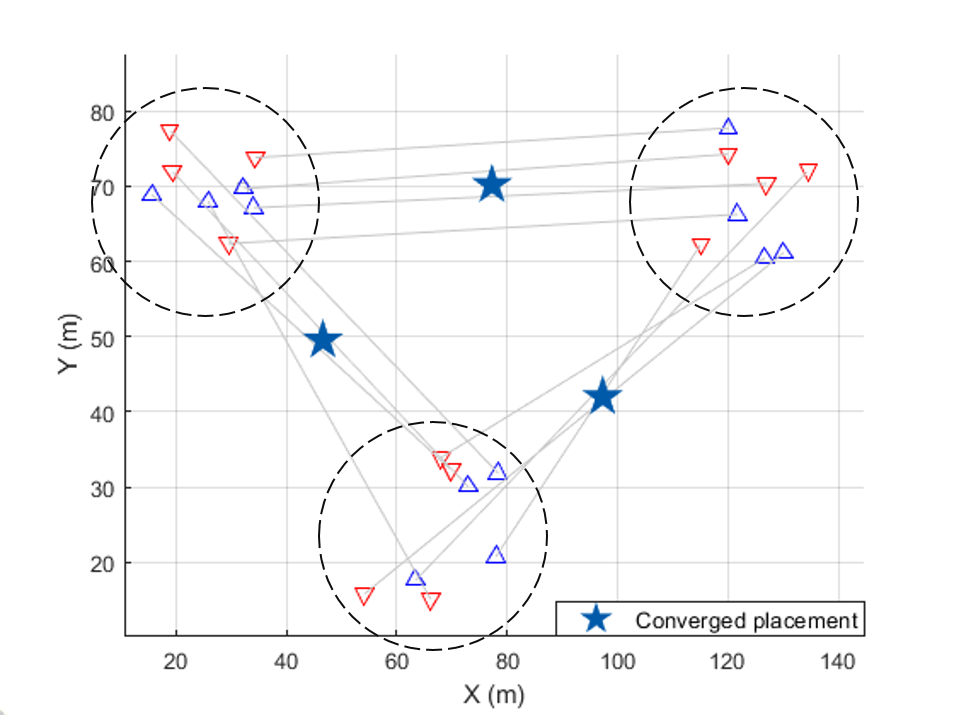}  
}     
\subfigure[Intra-cluster communication.] { 
\label{wc}
\includegraphics[width=3.1in]{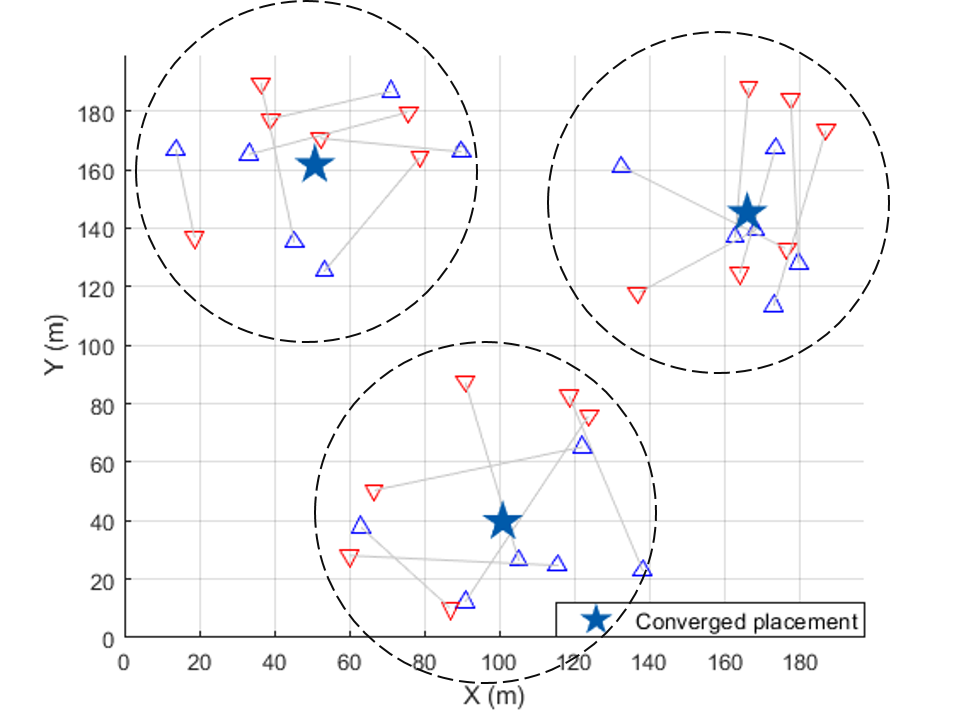}     
}    

\caption{Effects of ground-nodes spatial distribution.}  
\label{res7}     
\vspace{-22pt}
\end{figure}

Last, we show in Fig.~\ref{res7} the effects of the spatial distribution of ground nodes on the UAVs' placement. Different from the random spatial distribution considered in the previous subsections, we assume that the ground nodes form two types of clusters for communication, namely, the inter-cluster (see Fig.~\ref{bc}) versus intra-cluster (see Fig.~\ref{wc}) communications where the source and destination nodes are in different clusters and in the same cluster, respectively. It is observed that the converged UAVs' placement by our proposed IGS-BCD algorithm can adapt to both setups efficiently, i.e., the UAVs are located between far-apart clusters for the inter-cluster communication case (see Fig.~\ref{bc}) or within the clusters for the intra-cluster communication case (see Fig.~\ref{wc}), as expected.

\section{Conclusions}\label{clud}
\vspace{5pt}
In this paper, we studied the joint optimization of UAVs' 3D placement and bandwidth-and-power allocation in a multi-UAV relaying system. Under the elevation-angle dependent Rician fading UAV-ground channel model, an optimization problem was formulated to maximize the minimum {achievable expected rate} among multiple pairs of ground nodes. To solve this problem efficiently, we proposed a new \textit{IGS-BCD} algorithm by synergizing the advantages of both the GS and BCD methods. Moreover, we proposed a customized UAVs' placement initialization scheme for the proposed algorithm. 
Numerical results demonstrated the performance gains of the proposed algorithm as compared to various benchmark schemes including the conventional ones based on BCD or GS alone in terms of computational time as well as achievable rate. The proposed IGS-BCD algorithm is general and can also be applied to UAVs' placement optimization in other UAV-assisted communication systems.
\appendices

\section{proof for Lemma \ref{lem2}}
\vspace{10pt}
Using Lemma \ref{lem1}, it can be shown that $\tilde{g}(x, y)=(X+x) \log _{2}\left(1+{\gamma(Y+y)}/{(X+x)}\right)$ is concave w.r.t. $x>-X$ and $y>-Y$. Thus, we can upper-bound $\tilde g(x,y)$ by using the SCA technique. Specifically, for any given $x_0$ and $y_0$, we have $\tilde g(x,y)\leq \tilde g(x_0,y_0)+\tilde g_x(x_0,y_0)(x-x_0)+\tilde g_y(x_0,y_0)(y-y_0),\forall x,y$, where
\begin{align}
    &\tilde g_x(x_0,y_0) = \frac{-\gamma (Y+y_0)+\ln (1+\frac{\gamma (Y+y_0)}{X+x_0})((X+x_0)+\gamma (Y+y_0))}{((X+x_0)+\gamma (Y+y_0))\ln 2},\\
    &\tilde g_y(x_0,y_0) = \frac{(X+x_0) \gamma}{((X+x_0)+\gamma (Y+y_0))\ln 2}.
\end{align}
By setting $x_{0}=0$ and $y_{0}=0,$ we obtain
\begin{align}
    &(X+x)\log_2 \left(1+\dfrac{\gamma (Y+y)}{(X+x)}\right) \nonumber\\
    &\leq X\log_2 \left(1+\dfrac{\gamma Y}{X}\right)+\frac{-\gamma Y+\ln (1+\frac{\gamma Y}{X})(X+\gamma Y)}{(X+\gamma Y)\ln 2}x+\frac{X \gamma}{(X+\gamma Y)\ln 2}y.
\end{align}
By letting $\gamma = \frac{f(v_{m,k}^{(\mathrm{d})})\gamma_{0}}{({z_m^{2}+\|\boldsymbol{q}_{m}-\boldsymbol{u}^{(\rm d)}_{k}\|^{2}})^{\alpha/2}}$, $X = \hat a_{m, k}^{(\mathrm{d})}$, $x = a_{m, k}^{(\mathrm{d})}-\hat a_{m, k}^{(\mathrm{d})}$, $Y =\hat p_{m, k}^{(\mathrm{d})}$, and $y = p_{m, k}^{(\mathrm{d})}-\hat p_{m, k}^{(\mathrm{d})}$, we thus derive Lemma \ref{lem2}.
Similarly, $\hat{R}_{m,n,k}$, $\hat{\Omega}_{m,n,k}^{\rm{u b}}$, and $\hat{\Lambda}_{m,n,k}^{\rm{ ub}}$ can be defined in similar forms as $\hat{R}_{m, k}^{(\mathrm{d})}$, $\hat{\Psi}_{m, k}^{(\mathrm{d}),\rm{ub}}$, and $\hat{\Phi}_{m, k}^{(\mathrm{d}),\rm{ub}}$. The proof for Lemma \ref{lem2} is thus completed. 

\section{proof for Lemma \ref{lem3}}
\vspace{10pt}
This lemma can be proved by contradiction. To maximize the minimum achievable rate among all ground nodes in problem (P3.a), the equalities in \eqref{cons:infor_relax} and \eqref{cons:ObjOrig} for all destination nodes should hold, i.e., $\sum_{m \in \mathcal{M}} \tilde{R}_{m,k}^{(\rm d)} = \eta, \forall k \in \mathcal{K}$ and $\tilde R_{m, k}^{(\rm d)}+\sum_{n \in \mathcal{M}, n\neq m} R_{m,n,k}
$=$ \tilde R_{k, m}^{(\rm s)} +\sum_{n \in \mathcal{M}, n\neq m} R_{m,n,k}, \forall m\in\mathcal{M}$, $k\in\mathcal{K}$. Otherwise, we can always adjust the UAVs' 3D placement to make the equality hold without decreasing the objective value. 
For the constraint \eqref{cons:vmk}, if the inequality holds in the optimal solution to problem (P3.b), i.e., $v_{m,k}^{(\mathrm{d})}<\frac{z_m}{\sqrt{z_m^{2}+\|\boldsymbol{q}_{m}-\boldsymbol{u}^{(\rm d)}_{k}\|^{2}}}$, then we can always find another $\tilde v_{m,k}^{(\mathrm{d})}$ such that $\tilde v_{m,k}^{(\mathrm{d})} = \frac{z_m}{\sqrt{z_m^{2}+\|\boldsymbol{q}_{m}-\boldsymbol{u}^{(\rm d)}_{k}\|^{2}}}$. 
{\color{black}With the newly chosen $\tilde v_{m,k}^{(\mathrm{d})}>v_{m,k}^{(\mathrm{d})}$, the right-hand side of constraint \eqref{cons:dmkv} is decreased. As such, $r_{m,k}^{(\mathrm{d})}$ can be increased by letting $\tilde d_{m,k}^{(\mathrm{v})}=1+e^{-\left(B_{1}+B_{2} v_{m,k}^{(\mathrm{d})}\right)}$ in constraint \eqref{cons:dmkv}. The objective value, $\eta$, can be further improved by letting the equality in constraint \eqref{cons:ObjRef} hold, thus contradicting the assumption. Moreover, we can further increase $r_{m,k}^{(\mathrm{d})}$ by letting the equality in constraint \eqref{cons:dmkd} hold, because $r_{m,k}^{(\mathrm{d})}$ is  monotonically decreasing w.r.t. $\tilde{d}_{m, k}^{(\mathrm{d})}$, so that the objective value, $\eta$, can be further improved. The proof for the equality in constraint $\eqref{cons:vkm}$ is similar to that in \eqref{cons:vmk} and thus omitted for brevity.
Similarly, the equalities in constrains \eqref{cons:ObjRef} and \eqref{cons:InfoRef} should hold, i.e., $\sum_{m \in \mathcal{M}} r_{m, k}^{(\mathrm{d})}=\eta, \forall k \in \mathcal{K}$ and $ r_{m, k}^{(\rm d)}+\sum_{n \in \mathcal{M}, n\neq m} r_{m,n,k}
=\tilde R_{k, m}^{(\rm s)} +\sum_{n \in \mathcal{M}, n\neq m} R_{n,m,k}, \forall m\in\mathcal{M},k\in\mathcal{K}$. Otherwise, we can always adjust the locations of UAVs to change the value of $\tilde d_{m,k}^{\rm{(d)}}$ and $\tilde d_{m,n,k}$ so as to make the equality hold without decreasing the objective value. When the equality in constraint \eqref{cons:InfoRef} holds, we can always decrease $r_{m,n,k}$ to increase $r_{m, k}^{(\mathrm{d})}$ by increasing $\tilde{d}_{m, n, k}$, thus the equality in constraint \eqref{cons:dmnk} holds. The proof for Lemma \ref{lem3} is thus completed.}

\bibliographystyle{IEEEtran}

\end{document}